\documentclass[a4paper,10pt,twoside]{article}

\usepackage{amsmath,amssymb,amsthm}
\usepackage[ps,dvips,arc,frame,knot]{xy}
\usepackage[dvips]{graphicx}
\usepackage{psfrag}
\usepackage[T1]{fontenc}

\newcommand{\im}{\mathrm{i}}  
\newcommand{\defeq}{:=}

\newcommand{\N}{\mathbb{N}}
\newcommand{\Z}{\mathbb{Z}}

\newcommand{\R}{\mathbb{R}}
\newcommand{\C}{\mathbb{C}}
\newcommand{\tens}{\otimes}  

\newcommand{\xd}{\mathrm{d}}  
\newcommand{\xD}{\mathcal{D}}  

\newcommand{\lalg}[1]{\mathfrak{#1}}  

\DeclareMathOperator{\tr}{tr}

\DeclareMathOperator{\sdim}{Sdim}
\DeclareMathOperator{\qdim}{qdim}
\DeclareMathOperator{\qsdim}{qSdim}

\newcommand{\one}{\mathbf{1}}

\newcommand{\act}{\triangleright}

\renewcommand{\i}[1]{{}_{\scriptscriptstyle(#1)}}
\newcommand{\iu}[1]{{}_{\scriptscriptstyle(\underline #1)}}

\newcommand{\pfx}[1]{\mathcal{Z}_{\text{#1}}}
\newcommand{\pfhx}[1]{\hat{\mathcal{Z}}_{\text{#1}}}
\newcommand{\pfmx}[1]{\tilde{\mathcal{Z}}_{\text{#1}}}
\newcommand{\cd}{\mathcal{D}}
\newcommand{\cdh}{\hat{\mathcal{D}}}
\newcommand{\SU}{\mathrm{SU}}
\newcommand{\SL}{\mathrm{SL}}
\newcommand{\Sp}{\mathrm{Sp}}
\newcommand{\SO}{\mathrm{SO}}

\newcommand{\OSp}{\mathrm{OSp}}
\newcommand{\osp}{\lalg{osp}}
\newcommand{\su}{\lalg{su}}
\renewcommand{\sp}{\lalg{sp}}
\newcommand{\so}{\lalg{so}}
\newcommand{\iso}{\lalg{iso}}
\newcommand{\spin}{\lalg{spin}}

\theoremstyle{plain}
\newtheorem{prop}{Proposition}[section]

\theoremstyle{definition}

\newcommand{\rxy}[1]{{\begin{xy} 0;<1mm,0mm>:<0mm,1mm>::0;0,#1
\end{xy}}}

\def\be{\begin{equation}}
\def\ee{\end{equation}}
\def\bes{\begin{eqnarray}}
\def\ees{\end{eqnarray}}
\def\arr{\rightarrow}
\def\harr{\hookrightarrow}
\def\om{\omega}
\def\w{\wedge}
\def\la{\langle}
\def\ra{\rangle}
\def\f{\frac}

\newcommand{\Ref}[1]{(\ref{#1})}

\def\wtl{\widetilde}
\def\what{\widehat}
\def\inter{I^{j_1j_2}{}_{j_3}}

\title{Three-dimensional Quantum Supergravity\\
 and Supersymmetric Spin Foam Models}
\author{{\bf Etera R Livine}\thanks{email: livine@cpt.univ-mrs.fr},
{\bf Robert Oeckl}\thanks{email: oeckl@cpt.univ-mrs.fr} \vspace{2mm}\\ 
{\small Centre de Physique Th\'eorique,
Campus de Luminy, Case 907, } \\
{\small 13288 Marseille cedex 9, France }}
\date{\today}

\begin{document}

\maketitle

\begin{abstract}
We show how super BF theory in any dimension can be quantized as a
spin foam model, generalizing the situation for ordinary BF theory.
This is a first step toward quantizing supergravity theories.
We investigate in particular 3-dimensional $(p=1,q=1)$ 
supergravity which we quantize exactly. We obtain a
super-Ponzano-Regge model with gauge group $\OSp(1|2)$.
A main motivation
for our approach is the implementation of fermionic degrees of
freedom in spin foam models.
Indeed, we propose a description of the fermionic
degrees of freedom in our model.
Complementing the path integral approach we also discuss aspects of a
canonical quantization in the spirit of loop quantum gravity.
Finally, we comment on $2\! +\! 1$-dimensional quantum supergravity and
the inclusion of a cosmological constant.

\end{abstract}

\newpage
\tableofcontents
\newpage

\section{Introduction}

Spin foam models provide a non-perturbative approach to quantum
gravity \cite{baez:sf, dan:review, alej:review}.
They can be motivated both, as a rigorous way of performing a
covariant path integral quantization, and as emerging as histories
in the (canonical) loop approach to quantum gravity.
So far these approaches to quantum gravity have lived a largely
separate life from the main perturbative approach, namely string
theory. An obstacle to linking the two has been the fact that string
theory presumes
supersymmetry. In the spin foam context this would require
supergroups as gauge groups, something that so far had not been
realized. However, with the recent introduction of a new formalism
this limitation has been overcome \cite{Oe:qlgt}.\footnote{Note that it
is possible in principle to handle supergroups already by adapting the
formalism of Barrett and Westbury in dimension three
\cite{BaWe:invplm} or that of Crane, Kauffmann and Yetter in dimension
four \cite{CrKaYe:inv4}.
This would be rather laborious
however and has the crucial disadvantage that it does not generalize
to higher dimensions.}

We consider here the extension of the spin foam approach to the
quantization of supergravity theories. Apart from thus building a
bridge towards string theory this has important intrinsic
motivations.
Foremost this is the question of including matter degrees
of freedom into pure gravity spin foam models.
So far, various proposals have been made
both for fermions \cite{miko1,crane:matter} and for gauge fields
\cite{dan&hendryk,miko2}, but no consensus yet
has emerged on how to proceed.
Supergravity offers a new perspective at least for fermions. On
the one hand the
Lagrangian automatically contains fermions, on the other hand a
superfield formulation allows to ``hide'' these in the gauge field.
The quantization we
propose proceeds by taking advantage of the superfield
formulation. The generalization from fields to
superfields can be thought of as occurring on the level of the
gauge group, which becomes a supergroup. This allows to ``mimic''
the quantization of ordinary gravity. The resulting models then
appear as a guide as to how to encode fermions in more general
situations.

Our main focus here is three-dimensional supergravity, where we show
how the exact non-perturbative path integral quantization is performed. We
proceed in analogy to ordinary gravity in three dimensions. There,
using frame field and connection variables the Lagrangian takes the
form of BF theory. This is quantized through a discretization of
the path integral using methods of lattice gauge theory. The
result is independent of the discretization and provides thus an exact
quantization \cite{PoRe:limracah,ooguri}.

Using superfields, three-dimensional supergravity can be written in
the analogous way by employing superframe and superconnection
\cite{sugra}. The gauge group becomes $\OSp(1|2)$ in the $(p=1, q=1)$
version of supergravity. Using the formalism of circuit diagrams
\cite{Oe:qlgt} the quantization is then analogous to that of
gravity, but with the supergroup $\OSp(1|2)$ replacing $\SU(2)$.

While our quantization of supergravity is performed using the circuit
diagram formalism we also exhibit the conversion into the more
conventional spin foam
formalism. However, we note that a naive treatment of the resulting
expressions leads to potential ambiguities. This seems to be at the
origin of difficulties that have previously prevented the formulation
of supersymmetric models.

We also investigate the loop quantization of this supergravity theory,
which is believed to provide the canonical framework corresponding
to the path integral defined by the spin foam partition function.
This parallels the situation for ordinary (three-dimensional) gravity and
we use super-spin networks, using $\OSp(1|2)$ representations, instead
of the usual spin networks, using $\SU(2)$ representations.
Following the logic of loop quantum gravity, we
introduce geometric operators.
However, the usual length operator, constructed from the triad,
is not diagonal on the super-spin network
basis, and we need to introduce a super-length operator, constructed from
the super-triad. 
Then, using the relations between $\OSp(1|2)$ and $\SU(2)$ representations,
these states of quantum supergravity can be understood as superpositions
of pure (quantum) gravity states.

Based on properties of the representation theory of $\OSp(1|2)$ we
make a proposal for identifying the fermions in the spin
foam. Indeed, fermions are associated to edges of the spin foam and
the partition function can be understood as a sum over
gravity+fermions configurations, which are defined through the
identification of fermionic paths drawn on the spin foam.
This framework
for fermions is very similar to the loop quantum gravity point of view.
Nevertheless, in the present context, the fermionic degrees of freedom are
mixed together with the gravity degrees of freedom in such a way that
we can see the model both as a gravity+fermions system or
as sum over special superpositions -the supersymmetric ones-
of pure gravity configurations.

While we concentrate in this paper on supergravity without a cosmological
constant, its inclusion could be achieved similarly as with
gravity (see for example \cite{ooguri,signpb}).
That is, the cosmological constant should correspond to a
$q$-deformation of the gauge group, leading to a super-version
of the Turaev-Viro model. Indeed we present elements of such a construction
at the end,
using the quantum group $\OSp_q(1|2)$.

While we only consider supergravity in dimension three here, a crucial
property of our approach is its extendibility to higher
dimensions. This is again in parallel to the situation for gravity.
Namely, the Lagrangian of gravity in any dimension may be written as
that of BF theory plus a constraint \cite{fkp}. This suggests to quantize
gravity by first quantizing BF theory (which can be done exactly and
non-perturbatively) and then implementing the constraints at the
quantum level. This is indeed the key idea behind recent proposals for
spin foam models of quantum gravity \cite{bc}. Although, the focus has,
for obvious reasons, been on dimension four the same route can be
taken also in higher dimensions. Importantly, there are indications for
corresponding relations between supergravity theories and super BF
theory (see for example \cite{smolin:11dbf}).
This would also open the way to a
non-perturbative approach to M-theory via 11 dimensional quantum
supergravity. Crucially, the quantization of super BF
theory presented here works in any dimension.

In Section~\ref{sec:circdiag},
we introduce the diagrammatics necessary to define
supersymmetric spin foam models.
In Section~\ref{sec:3dgr}, we explain how
the quantization of BF theory leads to spin foam models and how to include
graded representations in the scheme. We illustrate this with two versions
of the Ponzano-Regge model:
the normal one based on $\su_2$ standard representation theory and
a graded one obtained by assigning parities to $\su_2$ representations
(i.e.~treating them as bosonic or fermionic depending on the spin).
We define precisely each model and  discuss
the relations between the two.
In Section~\ref{sec:sugra},
we show how three-dimensional supergravity can be written as a
(supersymmetric) BF theory. We also exhibit the representation theory of
$\osp(1|2)$ necessary for the quantization. Furthermore, we discuss the
canonical picture and aspects of its quantization.
In Section~\ref{sec:superPR},
we quantize three-dimensional supergravity as a Super-Ponzano-Regge model,
we study its formulation in term of supersymmetric 6j-symbols and
explore its geometry trying to disentangle the fermionic degrees of
freedom from the pure gravity degrees of freedom.
Finally we comment on the modifications necessary
for $2+1$-dimensional supergravity
and for including a cosmological constant.


\section{Circuit diagrams}
\label{sec:circdiag}

In order to efficiently represent and manipulate the state sum models
considered in this paper we use the formalism of
\emph{circuit diagrams} \cite{Oe:qlgt}. This is a diagrammatic language
to represent natural intertwiners of groups, supergroups or
quantum groups. While somewhat similar to spin network diagrams the
circuit diagrams have crucial advantages in supergroup or quantum group
settings.

The proper definitions are somewhat technical as they use the language
of category theory. However, we are here interested
only in diagrams arising in the context of representations of groups
and supergroups.\footnote{In terms of \cite{Oe:qlgt} these are circuit
diagrams for \emph{symmetric categories}. More precisely, they are
circuit diagrams for a category of $\Z_2$ graded objects.} The
present section gives an introduction to circuit
diagrams in this context. Furthermore, we explain their relation to
spin network diagrams.

\subsection{Diagrams for graded representations}

A circuit diagram consists of lines, called \emph{wires} and
rectangular boxes, called \emph{cables}. Each wire carries an
orientation, i.e., is equipped with an arrow. It also carries the
label of a representation. A wire can pass through
cables, entering at the top and leaving at the bottom. Wires may have
free ends, lined up on the top and bottom line of the diagram, but
nowhere else.

An (open) circuit diagram represents an intertwiner (i.e., an
invariant map) between
representations. Say the representations labeling the wires ending at
the top are $V_1,\dots, V_n$ and the representations labeling the wires
at the bottom are $W_1,\dots, W_m$. Then the diagram represents a map
$V_1\tens\cdots\tens V_n\to W_1\tens\cdots\tens W_m$. For wires with
an arrow pointing upward the respective representation is replaced
with its dual.

In order to evaluate a circuit diagram it is necessary to decompose it
into elementary diagrams. These are then pieced together horizontally
and vertically to yield the complete diagram. Horizontal composition
corresponds to taking the tensor product while vertical composition
corresponds to composition of maps.

\begin{figure}
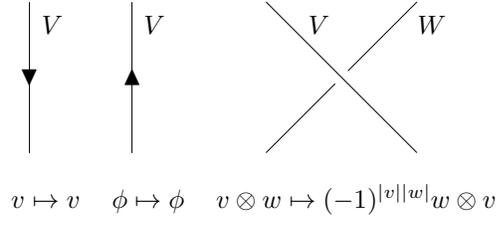
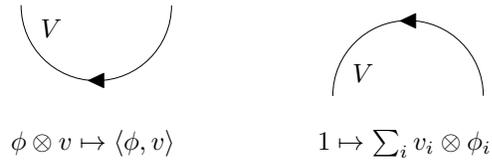
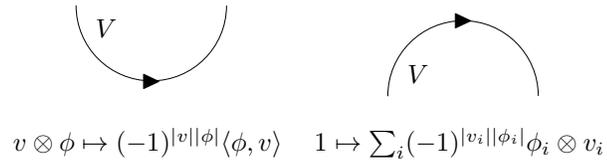

\begin{center}
\begin{tabular}{ccc}
\input{figures/fig_tan_id} & \input{figures/fig_tan_ids} &
\input{figures/fig_tan_xo}\\
\\
$v\mapsto v$ & $\phi\mapsto \phi$ &
$v\tens w\mapsto (-1)^{|v| |w|} w\tens v$
\end{tabular} \\
\vspace{0.5cm}
\begin{tabular}{cc}
\input{figures/fig_tan_archdl} & \input{figures/fig_tan_archul} \\
\\
$\phi\tens v\mapsto\langle \phi, v\rangle$ &
$1\mapsto\sum_i v_i\tens \phi_i$ \\
\\
\input{figures/fig_tan_archdr} & \input{figures/fig_tan_archur} \\
\\
$v\tens\phi\mapsto (-1)^{|v| |\phi|} \langle \phi, v\rangle$ &
$1\mapsto \sum_i (-1)^{|v_i| |\phi_i|} \phi_i\tens v_i$
\end{tabular}
\caption{\small Elementary diagrams and their corresponding
  intertwiners.}
\label{fig:elemtan}
\end{center}
\end{figure}

The elementary diagrams consisting of wire only are
listed in Figure~\ref{fig:elemtan}. Note that the lack of a wire
ending at the top or bottom of a
diagram means that one takes the trivial representation $\one$
which is identified with the complex numbers $\C$. Elements of $V$ and
$V^*$ are denoted $v$ and $\phi$ respectively. 
The pairing
$V^*\tens V\to\C$ between a representation $V$ and its dual $V^*$ is
denoted by $\phi\tens v\mapsto \langle \phi,v\rangle$. $\{v_i\}$
denotes a basis of $V$ and $\{\phi_i\}$ a dual basis of $V^{*}$.
For the crossing diagram, no arrows are drawn as the specified
intertwiner takes the same form for all arrow configurations.

\begin{figure}
\begin{center}
\input{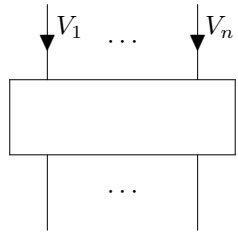}
\caption{\small A cable diagram with $n$ wires going through, labelled $V_1$
  to $V_n$.}
\label{fig:cable}
\end{center}
\end{figure}

The only other elementary diagram is the cable, see
Figure~\ref{fig:cable}. The intertwiner
\begin{equation}
T:V_1\tens\cdots\tens V_n\to V_1\tens\cdots\tens V_n
\label{eq:T}
\end{equation}
it corresponds
to is the projection of the tensor product representation onto
its trivial (i.e., invariant) subrepresentation. For representations
of a group $G$ we can express this using the Haar measure as
\begin{equation}
 T:v_1\tens\cdots\tens v_n \mapsto \int \xd g\, g\act
 v_1\tens\cdots\tens g\act v_n .
\label{eq:cabgroup}
\end{equation}
Here $g\act v$ denotes the action of $g$ on $v$.

Intuitively speaking, this is also what
is going on for supergroups. To give a rigorous definition in this
case, however, requires a Hopf algebraic setting. We give a rough
sketch of this here without going into further details.
For a group $G$ one can consider the Hopf algebra $H$ of algebraic
functions on $G$. The integral then becomes a map $\int: H\to
\C$. An action of the group becomes a coaction of $H$. This is a
map $V\to V\tens H$ written as $v\mapsto v\iu1\tens v\i2$. (The
subscript denote indices that are summed over.) Then the intertwiner
$T$ can be expressed as
\begin{equation}
 T:v_1\tens\cdots\tens v_n \mapsto v_1\iu1\tens\cdots\tens v_n\iu1
 \int v_1\i2 \cdots v_n\i2 .
\label{eq:cabhopf}
\end{equation}
This is equivalent to the expression (\ref{eq:cabgroup}). However, it
generalizes to the supergroup case. Namely, while a group
gives rise to a commutative Hopf algebra, a supergroup gives rise to a
graded commutative Hopf superalgebra $H$. The integral is again a map
$\int: H\to \C$ and (\ref{eq:cabhopf}) again defines the intertwiner
$T$ only with the difference that a factor
\[
(-1)^{\sum_{k=1}^n \sum_{j=2}^n |v_j\iu1| |v_k\i2|}
\]
has to be inserted to take care of the grading.

An important property of the circuit diagrams (of the type considered
here) is their invariance under
``combinatorial isotopy''.\footnote{Circuit diagrams for
  non-symmetric categories have weaker isotopy properties. This has
  the effect of placing constraints on the 
  possible space-time dimension of spin foam models with quantum gauge
  groups \cite{Oe:qlgt}.}
That is, two circuit diagrams evaluate to
the same intertwiner as long as they are combinatorially
identical. This means they need to be composed out of the same cables
and pieces of wire, connected in the same way, with the same labels,
arrow directions and wire endings. However, the spatial arrangement of
these components may be completely arbitrary otherwise (except for the
wire endings at the top and bottom of the diagram).

A closed circuit diagram represents an intertwiner $\one\to\one$ from
the trivial representation to itself, i.e., it is just a complex
number. The simplest such diagram is obviously a single closed
loop. Its value is easily computed from Figure~\ref{fig:elemtan} by
composing two opposing arches. Regardless of the arrow direction the
result for a representation $V$ is\footnote{Note that
  $|v_i|=|\phi_i|$ for a graded basis and its dual.}
\begin{equation}
 \sum_i (-1)^{|v_i| |\phi_i|}=\sum_i (-1)^{|v_i|}
 =\dim V_0 - \dim V_1 =\sdim V,
\label{eq:sdim}
\end{equation}
i.e., the \emph{superdimension}. Here, $V_0$ is the even and $V_1$ the
odd part of $V=V_0\oplus V_1$.

\subsection{The group case}
\label{sec:circgrp}

If we are dealing with a group the
circuit diagram formalism has a simple interpretation in terms of
integrals over matrix elements. Indeed the formula (\ref{eq:cabgroup})
for the cable when contracted with a dual basis of the
representations $V_1, \dots, V_n$ is just an integral over a product of
matrix elements. If the circuit diagram is closed, all matrix elements
are contracted to characters, i.e., traces of matrix elements.

Concretely, this implies the following rules for evaluating closed
circuit diagrams of non-graded group representations
\cite{Oe:renormdisc}. Attach a group element and a direction to each
cable. Associate a character with each closed loop of wire
corresponding to the representation of the wire. This character is
evaluated on the product of group elements associated with the cables
it traverses. The product of group elements is built from right to
left following the arrow direction on the wire. (The starting point is
irrelevant because of the trace property.) If the directions on cable
and wire are opposite the corresponding group element is replaced by
its inverse. The value of the circuit diagram is the value of
the complete expression with all group elements integrated over.

For graded representations of a group, similar evaluation rules hold
for closed circuit diagrams. First note that
formula (\ref{eq:cabgroup}) for the cable still holds. However,
factors of $(-1)$ (see Figure~\ref{fig:elemtan}) coming from the
grading have to be taken into account in assembling a circuit
diagram. As shown in \cite{Oe:spinstat} these various factors can be
subsumed into a simple extra rule familiar from Feynman diagrams with
fermions. Namely, the character of a fermionic representation (i.e.,
a fermionic loop) is to be
weighted with a factor of $(-1)$. Another way
to look at this is as a replacement of the character as a trace of
matrix elements by the supertrace.

In the general supergroup case, such simplified rules for the
evaluation of circuit diagrams no longer hold. One way to view this is
as being due to the noncommutativity of the Hopf superalgebra
of functions on the supergroup. This noncommutativity introduces
ordering ambiguities for the matrix elements in the integrals.

Besides the supergroup case, the formalism also allows the extension
to quantum groups. Then, more restrictive rules apply for the handling
of the diagrams and a full categorical interpretation is inevitable
\cite{Oe:qlgt}. We will not make use of this context here, but it
would be the relevant one for models with a cosmological constant, see
Section~\ref{sec:ospq}.

\subsection{Relation to spin networks}
\label{sec:circsn}

\begin{figure}
\begin{center}
\input{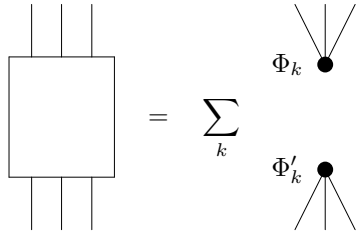}
\end{center}
\caption{Decomposition of a cable into spin network nodes.}
\label{fig:njdec}
\end{figure}

We now turn to the relation between circuit diagrams and spin network
diagrams. Essentially, a circuit diagram can be translated into a spin
network diagram, but not necessarily the other way round.
The expression of a circuit diagram in terms of spin networks is
effected by a decomposition of each cable into a pair of spin network
nodes, see Figure~\ref{fig:njdec}.

Recall that the cable represents
an intertwiner (\ref{eq:T}) that projects onto the trivial
subrepresentation. Decomposing this subrepresentation into
one-dimensional subspaces indexed by $k$ we can write
\begin{equation}
 T=\sum_k \Phi_k' \Phi_k
\label{eq:decsum}
\end{equation}
in terms of new intertwiners
$\Phi_k:V_1\tens\cdots\tens V_n\to\one$ and $\Phi_k':\one\to
V_1\tens\cdots\tens V_n$. These new intertwiners are now used to label
the nodes arising in cutting the cable (Figure~\ref{fig:njdec}).

When working with spin networks one normally makes once-and-for-all
choices of intertwiners to be associated with nodes for given
tensor products of representations. In particular, the intertwiners
are defined for arbitrary orientations of the legs of a node and
symmetric under reordering of the legs. (Indeed this is suggested by
the dot notation and familiar from the use of Feynman diagrams.)
Unfortunately, this has the effect that the nice isotopy properties of
the circuit diagrams are broken, in general. 
In other words, the value of a spin network diagram is not necessarily
the same for all combinatorial ways of assembling its elements (links
and nodes). This can mean that it is not well defined at all.

For non-graded group representations the problem can be and usually is
fixed. A clever choice of intertwiners ensures combinatorial
invariance of the spin network diagrams. For graded
representations of ordinary groups the situation is already more
tricky (with antisymmetries entering). For supergroups the
combinatorial invariance cannot be restored.
Indeed this seems to be at the root of the difficulties encountered in
previous efforts to define spin networks with supergroups and to define
supersymmetric spin foam models.


\section{3-Dimensional quantum gravity: two state sums}
\label{sec:3dgr}

In this section we review the path integral quantization of
three-dimensional gravity.
First, we review classical gravity in three dimensions and its formulation as
as a BF theory.
The path integral for the quantum theory, defining the dynamics
of three-dimensional loop quantum gravity, is defined as a spin foam model.
In this context, 
we start by considering the more general
setting of BF-theory
in arbitrary dimensions. Then specializing to gravity, we emphasize that
there are two versions of the Ponzano-Regge state sum. One comes
directly out of the quantization while the other results
from considering half-integer representations of $\SU(2)$ as
fermionic. This foreshadows aspects of the transition to
supersymmetric models.
Finally, we recall aspects of the canonical quantization.

\subsection{Classical gravity}

Let us start by describing the action for classical three-dimensional gravity
in a connection formalism. Let ${\cal M}$ be a three
dimensional manifold, the space-time.
The Lorentzian (resp. Euclidean) theory is formulated in terms of
two fields on ${\cal M}$:
a 1-form $\omega$ -the connection- valued in the
Lie algebra $\so(2,1)$ (resp. $\so(3)$)
and a triad field $e$ valued also in $\so(2,1)$ (resp. $\so(3)$).
Then the action is:
\be
S[\om,e]=\int_{{\cal M}} Tr \left(e\w F(\om) +\Lambda e\w e\w e\right)
\label{action3d}
\ee
where $F(w)$ is the curvature 2-form of $\om$ and
$\Lambda$ is the cosmological constant.
The equations of motion impose that the curvature scalar 
of the connection is equal to $\Lambda$
(the connection is flat when $\Lambda=0$)
and that the triad is compatible with the connection. The latter condition,
$d_\om e=0$, is the Gauss law imposing invariance under $\SU(1,1)$
(resp. $\SU(2)$) gauge transformations.
Therefore,
the theory is invariant under $\SO(2,1)$ (resp. $\SO(3)$) gauge transformation
and under diffeomorphisms.

It is then possible to formulate it as a gauge
theory of the Poincar{\'e} group \cite{witten:2+1}.
For this purpose, let us introduce the 1-form
\be
A_\mu=\om_\mu^i J_i +e_\mu^i P_i
\ee
where $J_i$ are the generators of $\so(2,1)$ (resp. $\so(3)$) and
$P_i$ the translation generators satisfying the following
commutation relations:
\be
\begin{array}{ccc}
\left[J_i,J_j\right] & = & \epsilon_{ij}^k J_k, \\
\left[J_i,P_j\right] & = & \epsilon_{ij}^k P_k, \\
\left[P_i,P_j\right] & = & \Lambda\epsilon_{ij}^k J_k.
\end{array}
\ee
$A$ is therefore valued in $\iso(2,1)$ (resp. $\iso(3)$) when $\Lambda=0$,
$\so(3,1)$ (resp. $\so(4)$) when $\Lambda> 0$ (de~Sitter case)
and $\so(2,2)$ (resp. $\so(3,1)$) when $\Lambda< 0$ (Anti-de~Sitter case):
$$
\begin{array}{|c|c|c|}
\hline
&{\rm Euclidean} & {\rm Lorentzian} \\
\hline 
\Lambda=0 & \iso(3) & \iso(2,1) \\
\hline 
\Lambda>0 & \so(4) & \so(3,1) \\
\hline 
\Lambda<0 & \so(3,1) & \so(2,2) \\
\hline
\end{array}
$$
Then the gravity action can be written as a Chern-Simons theory of $A$:
\be
S[A]=\int \la A\w dA\ra + \f{2}{3}\la A\w A\w A\ra
\ee
where $\la,\ra$ is an invariant quadratic form on the $(J_i,P_i)$
Lie algebra:
\be
\la J_i,J_j\ra=\la P_i,P_j\ra=0
\qquad
\la J_i,P_j\ra=\eta_{ij}.
\ee
In the Euclidean $\Lambda>0$ case, using $\so(4)\sim \spin(4)\sim
\su(2) \oplus \su(2)$, we can split the connection $A$ into selfdual and
antiselfdual parts $A_\pm$, which are 1-form valued in $\su(2)$. 
Indeed, let us introduce the generators
\be
J_i^\pm=\f{1}{2}\left(J_i\pm\f{1}{\sqrt{\Lambda}}P_i\right),
\ee
which satisfy the commutation relations $[J^\pm,J^\pm]=\epsilon J^\pm$
and $[J^+,J^-]=0$. Then the 1-form $A$ reads as
\be
A=\om^i J_i +e^i P_i
=A^{i+} J_i^+ +A^{i-} J_i^- \quad\textrm{with}\quad
A^{i\pm}=\om^i\pm\sqrt{\Lambda}e^i.
\ee
For $A^\pm$, we can introduce a Chern-Simons action
\be
S_\pm[A_\pm]=\int Tr_{su(2)}\left(A_\pm\w dA_\pm +
\f{2}{3} A_\pm\w A_\pm\w A_\pm\right),
\ee
which we can combine to get the initial action
\be
S[A]=\f{1}{4\sqrt{\Lambda}}(S_+-S_-).
\ee
We can also do the same for the Lorentzian $\Lambda<0$ case where
$\so(2,2)$ splits into $\sp(2) \oplus \sp(2)$ as a Lie algebra.

Finally, one can quantize the theory as a Chern-Simons theory or as
the sum of two Chern-Simons theories.
One can also quantize as a topological BF action
working directly on its path integral,
which leads to the Ponzano-Regge-Turaev-Viro model, the spin foam model
for 3-dimensional quantum gravity, which we review in the following
section \ref{sec:qbf}.
The link between the Turaev-Viro model and the Chern-Simons quantization
allows to track the role of the cosmological constant and relate it with
the quantum deformation of the $\SU(2)$ group used in the spin foam model.
One can also canonically quantize the theory in the spirit of loop
quantum gravity \cite{thiemann:2+1,2+1}, as explained in section
\ref{3dlqg}. It is understood to provide the canonical framework for the
path integral defined by the Ponzano-Regge model, though its link with
the Chern-Simons quantization is not yet clear.

\subsection{Quantization of BF theory}
\label{sec:qbf}

We briefly recall in this section the quantization of BF theory via
path integral and discretization (see for example \cite{baez:sf}).
We do this in terms of circuit diagrams on a cellular decomposition of the
underlying manifold
\cite{Oe:qlgt,GiOePe:diagtop,Oe:renormdisc}.

Consider the BF action
\begin{equation}
 S[A,B]=\int_M \tr(B\wedge F) ,
\label{eq:bfact}
\end{equation}
on a compact manifold $M$ of dimension $n$. Here $F$ is the curvature
2-form of a connection $A$ for a compact simple Lie group $G$. $B$ is
an $n-2$-form with values in the Lie algebra of $G$. $\tr$ is the
trace in the fundamental representation.

As a first step to evaluate the partition function we can formally
integrate out the $B$-field in the path integral as it appears
linearly in the exponential
\[
\pfx{BF}=\int \xD A \xD B\, e^{\im S[A,B]}=\int \xD A \,\delta(F).
\]
We remain with an integral over flat connections.

To make sense of the remaining path integral we discretize the
manifold $M$ as a CW-complex. In other words, we
divide it into open balls (``cells'') of dimension $n$. The spaces
left out we fill with open balls of dimension $n-1$, the spaces again
left with open balls of dimension $n-2$ etc.\footnote{Think of a foam
of soap bubbles: The interiors of the bubbles are the $n$-cells, the
walls where two bubbles meet are the $n-1$-cells, the lines where
several such walls meet are the $n-2$-cells etc.}
This is called a \emph{cellular decomposition} of $M$.

Now what we need is the 2-skeleton of the dual complex. This
means the following: Put into each $n$-cell a point, called
\emph{vertex}. Through each $n-1$-cell put a line, called an \emph{edge},
that connects the two vertices in the adjacent $n$-cells.
Finally, for each $n-2$-cell insert a surface, called a \emph{face},
that is bounded by the edges which pierce the adjacent $n-1$-cells.

Now we discretize the connection $A$ as in lattice gauge theory. That is,
we associate a group valued parallel transport $g_e$ with each edge $e$. We
then measure the curvature $F$ trough the holonomies, i.e., products of
parallel transports around the faces. The zero curvature condition
becomes the condition that the holonomies vanish, i.e., the product of
group elements around each face has to be the unit element. Finally,
the integral over connections becomes the Haar measure over group
elements for each edge.

The discretized version of BF theory thus has the following partition
function
\begin{equation}
 \pfx{BF}=\int \prod_e \xd g_e \prod_f \delta(g_1\cdots g_k).
\label{eq:bfdelta}
\end{equation}
Here $g_1,\dots, g_k$ are the group elements associated with the edges
bounding the face $f$.
Expanding the delta function in terms of characters of irreducible
representations
\begin{equation}
\delta(g)=\sum_V \dim V \chi_V(g)
\label{eq:delta}
\end{equation}
and reordering products and sums we obtain
\begin{equation}
\pfx{BF}=\sum_{V_f} \big(\prod_f \dim V_f \big) \int \prod_e \xd g_e
 \prod_f \chi_{V_f}(g_1\cdots g_k).
\label{eq:bfchar}
\end{equation}

\begin{figure}
\begin{center}
\includegraphics{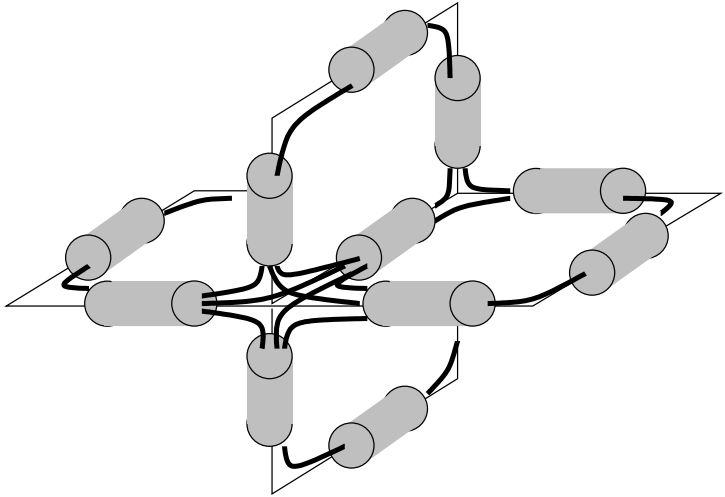}
\caption{A piece of the circuit diagram representing the partition function
  of BF theory. Labels and arrows for wires are omitted.}
\label{fig:pfdiag}
\end{center}
\end{figure}

We represent the partition function diagrammatically according to the
rules of Section~\ref{sec:circgrp}. Consider one big
circuit diagram embedded into the manifold and constructed as
follows. A closed wire is put into
each face along its bounding edges. It is labelled by the
representation $V_f$ associated with the face. A cable is put onto
each edge and the wires passing along this edge are routed through the
cable. This is depicted in Figure~\ref{fig:pfdiag}.
We denote the value of this labeled circuit diagram by $\cd(V_f)$.
The value of $\pfx{BF}$ is the value of $\cd(V_f)$ summed
over all assignments of irreducible representations to faces and
weighted by the product of the dimensions of the representations.

For $\pfx{BF}$ to be a good quantization of BF theory it should be
independent of the chosen cellular decomposition. This is almost the
case, except for an \emph{anomaly}. We define the
quantity\footnote{Note that this quantity is infinite, except for
  finite groups or certain quantum groups. Nevertheless, it
  makes sense in the diagrammatic calculus. Also, it presumably drops
  out of any physically measurable quantity.}
\begin{equation}
\kappa\defeq \sum_V (\dim V)^2,
\label{eq:kappa}
\end{equation}
where the sum runs over all irreducible representations $V$.
The partition function is modified to read
\[
 \pfmx{BF}\defeq \kappa^{p}\pfx{BF},
\]
where $p$ is a function of the number of cells of different
dimensions. The choice of $p$ can be viewed as a renormalization
problem \cite{Oe:renormdisc} and there are different ways to fix
it. For definiteness we shall set\footnote{Other choices (consistent
  with a TQFT interpretation) can be
  obtained by adding multiples of the Euler characteristic. A
  popular choice in dimension three is for example $p=-c_0$
  \cite{TuVi:inv3}.}
\[
 p\defeq -c_n+c_{n-1}-c_{n-2} ,
\]
where $c_k$ is the number of $k$-cells.
Let us point out that this is precisely the Euler characteristic
of the 2-complex.

The discretization independence of $\pfmx{BF}$ was proven in
dimensions 3 and 4 for simplicial decompositions in \cite{TuVi:inv3} and
\cite{Oog:toplat}. In dimension 3 it was extended to cellular
decompositions in \cite{GiOePe:diagtop}. In arbitrary dimensions the
proof was carried out in \cite{Oe:renormdisc}.

\subsection{Two versions of the Ponzano-Regge model}
\label{sec:pr}


We now turn back to the study of general relativity in three dimensions.
When $\Lambda=0$, the gravity action is
simply the BF-action (\ref{eq:bfact}) with
$G=\SO(3)$ and $A=\omega$, $B=e$. Thus, the path integral quantization
is given just by the partition function $\pfmx{BF}$ as described
above. That is, it can be expressed as a sum over circuit diagrams
embedded into a cellular decomposition of $M$. 
We shall denote the partition function in this context by $\pfx{PR}$
so that,
\begin{equation}
\pfx{PR} = \kappa^p \sum_{V_f} (\prod_f \dim V_f)\, \cd(V_f)
\label{eq:PRc}
\end{equation}
As described in
Section~\ref{sec:circsn} we may decompose the cables into pairs of
spin network nodes (Figure~\ref{fig:njdec}) by summing over
appropriate intertwiners $\Phi_e$ associated with edges $e$. The state
sum then acquires the form
\begin{equation}
\pfx{PR} = \kappa^p \sum_{V_f} (\prod_f \dim V_f) \sum_{\Phi_e} \prod_v
A_v(V_f,\Phi_e) ,
\label{eq:PoRe}
\end{equation}
which is more familiar for spin foam models. Here, $A_v$ are the
\emph{vertex amplitudes} that arise as the spin networks (recoupling
symbols) for the vertices.
Usually, a simplicial decomposition of $M$ is employed so that the
amplitudes $A_v$ all take the form of a $6j$-symbol. Furthermore, as
we shall assume in the following, we can use the group $G=\SU(2)$
instead of $\SO(3)$. This just means that we require $\omega$ to be a
spin connection, so as to allow for the possibility of fermionic
degrees of freedom.
The corresponding state sum is known as
the Ponzano-Regge model \cite{PoRe:limracah}.

To be more explicit, a simplicial decomposition is a decomposition
into tetrahedra. As each tetrahedron has six edges this means in terms
of the dual complex that each vertex $v$ is part of six faces
$f$. Thus there are six representation labels $j_1,\dots,j_6$
associated with each vertex. Furthermore, 2-cells are triangles and
thus have three 1-cells in their boundary. Dually this means that any
edge bounds three faces. As a consequence, there are exactly
three wires going through each cable. For the group $\SU(2)$ (or
$\SO(3)$) a tensor product of three irreducible representations has an
invariant subspace of dimension at most one. Thus, the sums
(\ref{eq:decsum}) arising in the decomposition of the cables
(Figure~\ref{fig:njdec}) have only one summand (or none) and we can
drop the intertwiner labels $\Phi_e$ in (\ref{eq:PoRe}). $A_v$ only
depends on $j_1,\dots,j_6$ and as a spin network it takes the form of
a tetrahedron. Its value is the $6j$-symbol and the
partition function takes the well known explicit form
\begin{equation}
\pfx{PR} = \kappa^p \sum_{j_f} (\prod_f (2j_f+1)) \prod_v
 \left\{\begin{matrix}j_1 & j_2 & j_3\\ j_4 & j_5 & j_6
 \end{matrix}\right\} .
\label{eq:PRexpl}
\end{equation}

A crucial point is that the representation of the partition function
$\pfx{PR}$ 
both in terms of circuit diagrams and in terms of recoupling symbols
does not directly refer to the gauge group $G$. Instead, only the
representations of $G$ appear. This allows us to modify
the representation structure in a way that is not induced by choosing
a different group $G$. For the group $\SU(2)$ there is the standard
choice considered so far. But there is also the choice of introducing
a grading on representations so that half-integer representations are
odd. Indeed, this is what the spin-statistics relation of quantum
field theory requires, if
the group $\SU(2)$ is to play the role of the covering group of the
rotations in 3-space. As this is the case in the
the Ponzano-Regge model we shall adopt this grading. Mathematically
one might view this as choosing a
``different'' version $\SU'(2)$ of $\SU(2)$ \cite{Oe:spinstat}. (This
also fits nicely with a categorical formulation of quantum field
theory \cite{Oe:bqft}.)

The implementation is completely straightforward in the circuit diagram
representation of the partition function. The change is captured
completely by the alteration of the rules for elementary diagrams
according to Figure~\ref{fig:elemtan}. The cumulative effect of these
modification is straightforward to evaluate following the rules of
Section~\ref{sec:circgrp} for graded representations of
groups. Namely, each closed wire labelled with an odd representation
acquires a factor of $-1$. Now, each representation label $V_f$ appears
exactly twice in the circuit diagram representation, both originating
from the expansion of the delta function (\ref{eq:delta}).
The first occurrence is in the factor $\dim V_f$, while the second is
in the character $\chi_{V_f}$ represented as the wire going round the
face $f$. The former, however, is really also a diagram, namely a
closed loop of wire. It only happens to equal the dimension of the
representation in the context of the derivation. In the more general graded
case it is the superdimension (\ref{eq:sdim}). For the spin $j$
representation of $\SU(2)$ this is $(-1)^{2j}(2j+1)$. Thus, we have
two factors of $-1$ for each odd representation that cancel. This
proves that the graded Ponzano-Regge model and the usual one agree in
their partition function.
The agreement between the two models ends, however, if we insert
observables that probe the signs or if we include boundaries.

To emphasize the modification we write the graded Ponzano-Regge model
as 
\begin{equation}
\pfhx{PR} = \hat{\kappa}^p \sum_{V_f} (\prod_f \sdim V_f)\,
 \cdh(V_f) . 
\label{eq:PRcg}
\end{equation}
This also exhibits another subtlety. Namely, in the definition
(\ref{eq:kappa}) of
$\kappa$ the dimension has to be replaced by the superdimension (hence
the notation $\hat{\kappa}$). This is again because the proper
definition of $\kappa$ is really as a sum over squares of loop diagrams.
For the present model this doesn't make
any difference as ordinary dimension and superdimension differ at most
by a sign for irreducible representations.

The spin foam representation for the graded Ponzano-Regge model is
obtained as usual from the circuit diagram representation. Now, the
modified rules for elementary diagrams (Figure~\ref{fig:elemtan}) have
to be used. This leads (as already described) to a replacement of the
dimension factors in (\ref{eq:PoRe}) by the superdimension factors and
to a modification of the $6j$-symbols $A_v$. In this formulation the
equality of the partition function between the two models is less easy
to recognize. Indeed it seems that the two versions of the
Ponzano-Regge model have been occasionally confused in the
literature.

We shall regard the graded model $\pfhx{PR}$ as
the physical Ponzano-Regge model. Indeed this will turn out to be
relevant in the supersymmetric setting.

Writing down a partition function is not sufficient to define
a quantum theory. Indeed, in the present case of three-dimensional gravity
the quantum theory is really defined by extending the Ponzano-Regge model to
a topological quantum field theory. This means
considering manifolds with boundary. State spaces are associated to
boundaries and the Ponzano-Regge state sum becomes a transition amplitude
between these state spaces, see e.g.\ \cite{ooguri}.
We describe this in detail in
Section~\ref{sec:boundary} in the more general supersymmetric case.

\subsection{Canonical framework: Loop Quantum Gravity}
\label{3dlqg}

Complementary to the path integral approach, it is also interesting to
study the canonical quantization of the BF theory formulation.
This leads to three-dimensional {\it loop quantum gravity}, on which
the reader can
can find details in the Euclidean case in \cite{thiemann:2+1}
and in the Lorentzian context in \cite{2+1}.

Forgetting the cosmological constant,
canonically analyzing action \Ref{action3d}, the pairs of conjugate
canonical variables are made of (the space components of) the triad $e$
and the connection $\om$. The Hamiltonian is entirely made of constraints.
The first one, $d_\om e=0$, is the Gauss law imposing gauge invariance
under $\SU(2)$ or $\SU(1,1)$\footnote{
$\SU(1,1)\sim\SL(2,\R)\sim\Sp(2)$ is the double
cover of $SO(2,1)$ but it is not its universal cover.
The first order formulation of gravity uses the triad and therefore
needs spinor indices provided by $SU(2)$ in the Euclidean case and
$SU(1,1)$ in the Lorentzian case.
Moreover, representations
of $\SU(1,1)$ are associated a parity $\epsilon=\pm 1$ (whether there are also
representations of $SO(2,1)$ or not), whereas representations of the universal
cover would be labelled by a continuous ''parity'' $\epsilon\in[0,1]$.
Therefore using $SU(1,1)$ to formulate a Lorentzian quantum theory
leads to structures very similar to those of the Euclidean theory.}.
The second one $F=0$ can be decomposed
into some space components generating space diffeomorphisms and
a time component -the Hamiltonian constraint-
governing the dynamics of the theory.
Then one can choose ($\SU(2)$ or $\SU(1,1)$) spin networks
(or more precisely equivalence classes of spin networks under
space diffeomorphisms) as partial observables\footnotemark, which we use
as kinematical states of the quantum theory. The last step is to study their
dynamics, i.e.~find the states satisfying the Hamiltonian constraint.
At the end of the process, one should recover gauge invariant states 
satisfying the $F=0$ flatness constraint.

\footnotetext{True Dirac observables should commute with all the
  constraints. Partial observables do not. Here they are quantities
  which are chosen to be gauge invariant (under the Gauss law) and
  diffeomorphism invariant. But the Hamiltonian constraint will act
non-trivially on them, i.e.\ they have a non-trivial time evolution. That
  is why we call them "kinematical states". Then we would like to
  project this space of kinematical states onto physical states
  satisfying also the Hamiltonian constraints. The reader can find a
  discussion on the notion of partial observables and their use
  in \cite{partialobs}.}

In the framework of loop quantization, there are ambiguities on the
implementation of the dynamics and we do not know yet a ``good choice''
of quantum Hamiltonian constraint.
It turns out that spin foams allow to adress this issue: instead of
canonically quantizing the Hamiltonian constraint, we construct the
path integral for the theory, which provides us with a projector onto
physical states solution of the Hamiltonian constraint.
Indeed spin foams appear as histories of spin networks and spin foam models
might be viewed as
loop gravity models taking into account the dynamics.
More precisely, when one looks at the boundary (or a two-dimensional slice)
of a 2-complex (a spin foam) in the Ponzano-Regge model, one gets
$\SU(2)$ spin networks (or $\SU(1,1)$ spin networks in
its Lorentzian theory). 
From this point of view, the Ponzano-Regge model appears as the space-time
version or path integral formulation for three-dimensional loop
quantum gravity.
Then one can check that the Ponzano-Regge partition function defines
a projector (in the space of the boundary states, see the comments
at the end of Section~\ref{sec:pr})
onto the $F=0$ sector and thus projects onto physical states of
the canonical theory \cite{ooguri}.

In the case of a cosmological constant, only the Hamiltonian constraint
is modified, and the Turaev-Viro model (deformed Ponzano-Regge model)
still allows to project onto states satisfying the constraints \cite{ooguri}.


\section{3-Dimensional classical supergravity}
\label{sec:sugra}

\subsection{Lagrangian formulation}
\label{csugra}

One can extend the formulation of three-dimensional gravity,
described in the previous section,
to the supersymmetric case of supergravity.
This is achieved in the
Lorentzian Anti-de~Sitter case by extending the symmetry group
$\Sp(2)\times \Sp(2)$ to $\OSp(p|2)\times \OSp(q|2)$. These $(p,q)$
type AdS supergravities were introduced by Ach{\'u}carro and
Townsend \cite{sugra}. The same procedure works in
the Euclidean de~Sitter case and we extend $\SU(2)\times \SU(2)$
to $\OSp_E(p|2)\times \OSp_E(q|2)$ (where the subscript $E$ means that
the bosonic part of the supergroup is
$O(p)\times \SU(2)$ and not $O(p)\times \Sp(2)$).

In the present paper, we are interested in the simplest supergravity case
$(p=1,q=1)$ with a $\OSp_E(1|2)\times \OSp_E(1|2)$ supersymmetry.
Using $\su(2)$ spinor indices $A,B,..=\pm$ and the notation $(AB)$
(resp. $[AB]$) for symmetrising (resp. antisymmetrising) the couple of
indices $AB$, the superalgebra $\osp_E(1|2)$ reads
\be
\begin{array}{ccl}
\left[J_{AB},J^{CD}\right]&=&\delta_{[A}^{[C}J_{B]}^{D]}, \\
\left\{Q_A,Q_B\right\}&=& J_{AB}, \\
\left[J_{AB},Q_C\right]&=&\epsilon_{C(A}Q_{B)}.
\end{array}
\ee
Then, one can introduce the supertriad
\be
{\cal E}=e^{AB}J_{AB}+\phi^AQ_A,
\ee
and a 1-form ${\cal A}$ valued in $\osp_E(1|2)$ -the superconnection:
\be
{\cal A}=\om^{AB}J_{AB}+\psi^AQ_A,
\ee
whose curvature is defined as ${\cal F}=\wtl{F}^{AB}J_{AB}+F^AQ_A$
with components
\bes
\wtl{F}^{AB}&=&F^{AB}(\om)+\psi^A\w \psi^B, \nonumber \\
F^A&=&d_\om\psi^A=d\psi^A+\om^A_B\w \psi^B.
\ees
The supersymmetric BF action for supergravity then reads
\be
S_{sugra}[{\cal E},{\cal A}]=
\int_{\cal M}STr\left({\cal E}\w {\cal F}({\cal A})
+\Lambda {\cal E}\w {\cal E}\w {\cal E}\right).
\label{bfsugra}
\ee
The equations of motion are ${\cal F}+\Lambda {\cal E}\w {\cal E}=0$
imposing that the curvature scalar is $\Lambda$ and the super Gauss law
$d_{{\cal A}}{\cal E}=0$ imposing the ordinary $\SU(2)$ gauge invariance
and the (left-handed) supersymmetric constraint.

Introducing the selfdual and antiselfdual components makes the 
$\osp_E(1|2)\oplus \osp_E(1|2)$ symmetry explicit.
Let us define
\bes
A^{AB}_\pm&=&\om^{AB}\pm\sqrt{\Lambda}e^{AB}, \nonumber\\
\psi^A_\pm&=&\psi^A\pm\Lambda^{1/4}\phi^A.
\ees
Then the supergravity theory can be expressed as the sum
of two Chern-Simons theories on  the 1-forms valued in $\osp_E(1|2)$,
$A^{AB}_+J_{AB}+\psi^A_+Q_A$
and $A^{AB}_-J_{AB}+\psi^A_-Q_A$, with the two fermionic fields
$\psi_\pm$:
\be
S_{sugra}=\f{1}{\sqrt{\Lambda}}\int_{\cal M}
I_{CS}(A^{AB}_+J_{AB}+\psi^A_+Q_A)-I_{CS}(A^{AB}_-J_{AB}+\psi^A_-Q_A)
\ee
with
\be
I_{CS}(A_\pm,\psi_\pm)=
STr_{osp_E(1|2)}\left(
A_\pm\w dA_\pm +\f{2}{3}A_\pm\w A_\pm\w A_\pm +
\psi_\pm\w d_{A_\pm} \psi_\pm
\right),
\ee
which is the form in which the supergravity theories are introduced
in \cite{sugra}.

Using this formalism, we are going to quantize
three-dimensional supergravity
as a super Ponzano-Regge model with symmetry group $\OSp_E(1|2)$
instead of $\SU(2)$. This state sum model will be a topological theory
based on the representations of the superalgebra $\osp_E(1|2)$
and, in the framework of spin foam models, it
should implement a (discrete) path integral for 
gravity plus fermions in three dimensions.


\subsection{Representation theory of $\lalg{osp}_E(1|2)$}
\label{sec:reposp}

Instead of using the supergroup $\OSp_E(1|2)$ directly we shall use
the super Lie algebra $\lalg{osp}_E(1|2)$ to obtain its
representations. This is technically easier and allows the recurse to
well established results in the literature. We review these in the
present section.

\subsubsection{Generators and action}

To study the representations of $\lalg{osp}_E(1|2)$, it is useful to write
the algebra as follows\footnotemark:
\bes
[J_3,J_\pm]=\pm J_\pm  \qquad [J_+,J_-]=2J_3 \nonumber\\
\left[J_3,Q_\pm\right]=\pm\f{1}{2}Q_\pm \qquad [J_\pm,Q_\pm]=0 
\qquad [J_\pm,Q_\mp]= Q_\pm  \nonumber \\
\{Q_\pm,Q_\pm\}=\pm\f{1}{2}J_\pm \qquad   \{Q_\mp,Q_\pm\}=-\f{1}{2}J_3
\label{commrelations}
\ees
\footnotetext{In terms of the $\su(2)$ generators $J_{1,2,3}$, the
algebra reads
$$
[J_i,Q_A]=(\gamma_i)_A^BQ_B \qquad
\{Q_A,Q_B\}=(\gamma_i)_{AB}J_i=(\epsilon\gamma_i)_A^BJ_i
$$
where the $\gamma_i$ are the Pauli matrices and $\epsilon$
the antisymmetric matrix:
$$
(\gamma_1)_A^B=\f{1}{2}
\left(\begin{array}{cc}
 0 &1 \\ 1 & 0
\end{array}\right) \qquad
\gamma_2=\f{i}{2}
\left(\begin{array}{cc}
0 & -1 \\ 1 &0
\end{array}\right) \qquad
\gamma_3=\f{1}{2}
\left(\begin{array}{cc}
1 & 0 \\ 0 & -1
\end{array}\right) \qquad
\epsilon=
\left(\begin{array}{cc}
0 & 1 \\ -1  & 0
\end{array}\right).
$$}
The Casimir operator is $C=J^iJ^i+Q_+Q_--Q_-Q_+$. The
representations \cite{susyrep} are labelled by half-integer $j$
and are made out of pairs of $\su(2)$ representations carrying spin
$j$ and $j-1/2$:
\be
R^j=V^j\oplus V^{j-\f{1}{2}}
\qquad\forall j\ge \f{1}{2}.
\ee
We call $j$ the spin of the $\lalg{osp}(1|2)$ representations, $k=j,j-1/2$
the two isospins, and we label $|j,k,m\ra$
the vectors of a basis of $R^j$ using the usual notations for $\su(2)$.
More precisely, the action of the generators in the $j$ representations is:
\bes
J_3|j,j,m\ra&=&m|j,j,m\ra,\nonumber \\
J_3|j,j-\f{1}{2},m\ra&=&m|j,j-\f{1}{2},m\ra,\nonumber \\
J_\pm|j,j,m\ra&=&
\left((j\mp m)(j\pm m +1)\right)^{1/2}|j,j,m\pm1\ra,\nonumber\\
J_\pm|j,j-\f{1}{2},m\ra&=&
\left((j-\f{1}{2}\mp m)(j+\f{1}{2}\pm m)\right)^{1/2}
|j,j-\f{1}{2},m\pm1\ra,\nonumber\\
Q_\pm|j,j,m\ra&=&\mp(j\mp m)^{1/2}
|j,j-\f{1}{2},m\pm\f{1}{2}\ra,\nonumber\\
Q_\pm|j,j-\f{1}{2},m\ra&=&-\f{1}{2}(j+\f{1}{2}\pm m)^{1/2}
|j,j,m\pm\f{1}{2}\ra.
\ees
In fact, we have two levels $V^j$ and $V^{j-1/2}$. The $\su(2)$ generators
$J_{3,\pm}$ act as usual on each level. Then the supersymmetric generators
allow to go from one level to the other.
\begin{figure}[t]
\begin{center}
\psfrag{j}{$J_\pm$}
\psfrag{q}{$Q_\pm$}
\psfrag{m}{$m$}
\psfrag{mp}{$m+1$}
\psfrag{mm}{$m-1$}
\psfrag{mpd}{$m+\f{1}{2}$}
\psfrag{mmd}{$m-\f{1}{2}$}
\psfrag{vj}{$V^j$}
\psfrag{vj1}{$V^{j-\f{1}{2}}$}
\includegraphics[width=6.5cm]{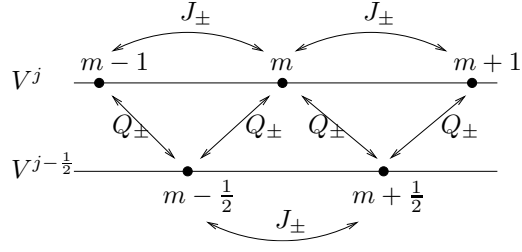}
\end{center}
\caption{Structure of the $R^j$ representation of $\lalg{osp}(1|2)$.}
\end{figure}
The Casimir of the $R^j$ representation is
\be
C_j=j\left(j+ \f{1}{2}\right).
\ee
So the fermions contribution is $-\f{j}{2}$ on the subspace $V^j\harr R^j$
and $\f{1}{2}\left(j+\f{1}{2}\right)$ on $V^{j-1/2}$.

\subsubsection{Parity}

Due to the ``anti-commuting'' properties of the $Q$ generators, we cannot
define a simple star operator and talk about unitary representations.
We need to introduce the parity (bosonic or fermionic) of the vectors
of the representations, use a grade star operation and talk about grade-star
representations \cite{susyrep}.

In a given representation $R^j$, we introduce a parity $\lambda=0,1$
which means that
 the vectors $|j,j,m,\lambda\ra$ of $V^j$ have parity $\lambda$
(0 for bosonic and 1 for fermionic) and that the vectors
$|j,j-1/2,m,\lambda+1\ra$ of $V^{j-1/2}$
have parity $\lambda+1$.

The grade star operation 
${}^\dagger$ will take into account the degree of the operators/vectors
(whether it is bosonic or fermionic).
More precisely, given an operator $A$ of degree $\alpha$,
$A^\dagger$ is defined as:
\be
\la A^\dagger\psi|\varphi\ra=(-1)^{\alpha\xi}\la \psi|A\varphi\ra
\quad\textrm{with }\xi\textrm{ the degree of }\psi.
\ee

With this definition we can require the hermiticity relations:
\be
J_i^\dagger=J_i \quad Q^\dagger_+=\mp Q_- \quad
Q^\dagger_-=\pm Q_+.
\label{Qhermicity}
\ee
Hermiticity of the $J$ generators imposes that
$$
\begin{array}{rcc}
\la j,j,m,\lambda|j,j,m',\lambda\ra&=&g\delta_{m,m'}\\
\la j,j-\f{1}{2},m,\lambda+1|j,j-\f{1}{2},m',\lambda+1\ra&=&h\delta_{m,m'},
\end{array}
$$
where $g,h$ are just signs.
Hermiticity of the $Q$ generators then implies $h=\pm(-1)^\lambda g$.
Finally, we have a grade star representation with a positive definite scalar
product iff $g=h$, i.e.\ $\pm(-1)^\lambda=1$
(with $\pm$ defined in \Ref{Qhermicity}).

So far every irreducible representation $R^j$ appears twice, once for
each choice of parity. However, we might wish to allow only one
parity to occur. Indeed, if the sub Lie algebra
$\lalg{su}_2\subset \lalg{osp}(1|2)$ plays the physical role of
(infinitesimal) rotations the representations must obey the
spin-statistics relation of quantum field theory. That is, the $V_j$
should be even or odd
depending on whether $j$ is integer or not. In the Hopf algebraic
approach to supergroups it is actually possible to encode this
restriction in the supergroup \cite{Oe:qgeosusy}. There is an ordinary
(not super but noncommutative) Hopf algebra that encodes
$\OSp(1|2)$. This has representations of both parities. Then there are
restricted Hopf algebras which allow for each irreducible
representation to occur with only one parity. The version we consider
here is called $\OSp'(1|2)$ in \cite{Oe:qgeosusy}.\footnote{There is
  another subtlety which has not been sufficiently appreciated in the
  literature. $\OSp_E(1|2)$ is supposed to have a bosonic subgroup
  $\SU(2)\times O(1)$ with $O(1)=\Z_2$ (and this is indeed the case of
  $\OSp(1|2)$ in \cite{Oe:qgeosusy}). However the representation
  theory of $\lalg{osp}(1|2)$ as considered here does not allow for an
  appropriate $\Z_2$ action. This means that it really corresponds to
  some ``truncated'' version of $\OSp(1|2)$. Nevertheless we shall be
  content with this truncation since it is self-consistent and the bosonic
  subgroup $\SU(2)$ is properly contained.\label{fn:osp}}
We shall continue to pursue the super Lie algebraic point of view, but
our remark ensures that the restriction of the parity is consistent.
Technically
speaking it means that we have a ``good'' category of representations.

The implications of the restriction in parity are as follows.
For $j\in\N$,
$Q_+^\dagger=-Q_-$ and $Q_-^\dagger=Q_+$ on $R^j$
and, for $j\in\N+1/2$, $Q_+^\dagger=Q_-$ and $Q_-^\dagger=-Q_+$ on $R^j$.
The superdimension of the representation is given by the
supertrace of the identity and yields:
\be
\sdim(R^j)=(-1)^{2j}\dim(V^j)-(-1)^{2j}\dim(V^{j-1/2})
=(-1)^{2j}.
\ee

\subsubsection{Tensor products}

Let us now describe the tensor products of representations $R^j$.
Taking into account that $R^j=V^j\oplus V^{j-1/2}$ and the recoupling
theory of $\su(2)$ representations, we get:
\be
R^{j_1} \otimes R^{j_2} =\bigoplus_{|j_1-j_2|\le j\le (j_1+j_2)}R^j,
\ee
where the sum over $j$ goes by {\it half-integer} steps.
Therefore, we see that the situation is very similar to the $\su(2)$ case
with a {\it triangular inequality} on the representations and the only
difference is that the representation $j$ takes all half-integer values
and not only integer steps as in the representation theory of $\su(2)$.
As an example, we now have
$$
R^{\f{1}{2}}\otimes R^{\f{1}{2}}=
R^0 \oplus R^{\f{1}{2}}\oplus R^1
$$
instead of $V^{1/2}\otimes V^{1/2}= V^0\oplus V^1$.
Then the intertwiner
$I^{j_1j_2}{}_{j_3}:R^{j_1} \otimes R^{j_2} \arr R^{j_3}$
is unique (up to normalization)
and we can deduce the corresponding
(supersymmetric) Clebsh-Gordan coefficients expressing the change of basis
between the vectors $|j_1 k_1 m_1\ra \otimes |j_2 k_2 m_2\ra$ and 
$|(j_1,j_2) j_3 k_3 m_3\ra$.

The simplest way to get the recoupling coefficients is to work with the isospin
decomposition $R^j=V^j\oplus V^{j-\f{1}{2}}$ and use the known
$\su(2)$ Clebsh-Gordan coefficients. Then the restriction of the intertwiner
of the $k_1=j_1,k_2=j_2$ isospins
$\inter :V^{j_1}\otimes V^{j_2}\arr V^{j_3}\oplus V^{j_3-\f{1}{2}}$
is invariant under $SU(2)$. Therefore either the image is $V^{j_3}$ or
$V^{j_3-\f{1}{2}}$ depending whether $j_1+j_2+j_3$ is an integer or not.

This way,  we see that they are two different situations:
\begin{itemize}
\item $j_1+j_2+j_3\in\N$: \\
The only non-vanishing components are
$V^{j_1}\otimes V^{j_2}\arr V^{j_3}$, which is the usual $\su(2)$ intertwiner
and its supersymmetric
counterparts $V^{j_1-\f{1}{2}}\otimes V^{j_2}\arr V^{j_3-\f{1}{2}}$
and $V^{j_1}\otimes V^{j_2-\f{1}{2}}\arr V^{j_3-\f{1}{2}}$.
Each of these intertwiners are given up to a factor by the $\su(2)$
Clebsh-Gordan coefficients. Then their relative normalization is fixed
by invariance of $\inter$ under the supersymmetric generators $Q_\pm$,
so that $\inter$ is unique up to an overall factor.
One can find the explicit expressions of the relative factors
in \cite{susyrep}.

\item $j_1+j_2+j_3\in\N+\f{1}{2}$:\\
The only non-vanishing components are
$V^{j_1-\f{1}{2}}\otimes V^{j_2-\f{1}{2}}\arr V^{j_3-\f{1}{2}}$,
and its supersymmetric
counterparts $V^{j_1}\otimes V^{j_2}\arr V^{j_3-\f{1}{2}}$,
$V^{j_1-\f{1}{2}}\otimes V^{j_2}\arr V^{j_3}$
and $V^{j_1}\otimes V^{j_2-\f{1}{2}}\arr V^{j_3}$.
Once again, each of these intertwiners is given  up to a factor by
the $\su(2)$
Clebsh-Gordan coefficient and their relative normalization is fixed by
the invariance under the supersymmetry generators. $\inter$ is then unique
up to an overall normalization.
\end{itemize}
\begin{figure}[t]
\begin{center}
\psfrag{eq}{$=$}
\psfrag{r1}{$R^{j_1}$}
\psfrag{r2}{$R^{j_2}$}
\psfrag{r3}{$R^{j_3}$}
\psfrag{int}{$j_1+j_2+j_3\in\N$}
\psfrag{halfint}{$j_1+j_2+j_3\in\N+\f{1}{2}$}
\includegraphics[width=9.5cm]{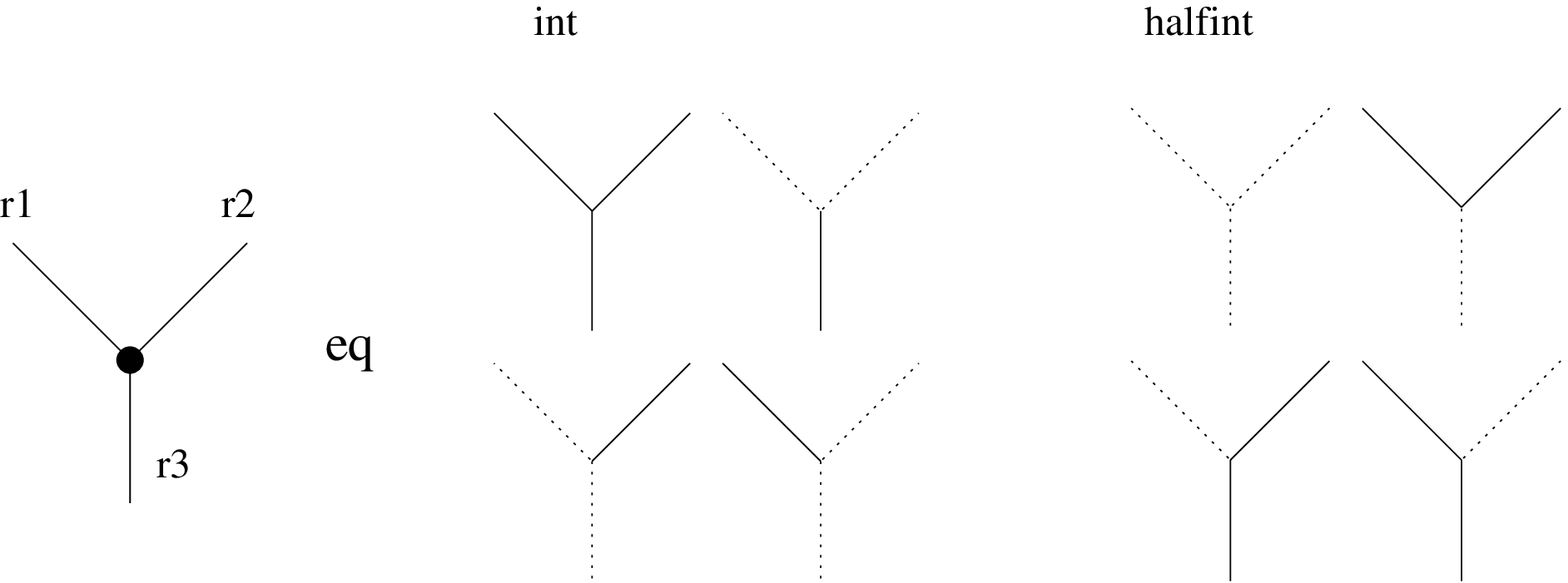}
\end{center}
\caption{\small Intertwiner $\inter:R^{j_1} \otimes R^{j_2} \arr R^{j_3}$
  invariant under $\lalg{osp}(1|2)$:
the decomposition on isospins $V^{j_i}$ (solid lines)
and $V^{j_i-1/2}$ (dotted lines) in the two cases
$j_1+j_2+j_3\in\N$ and $j_1+j_2+j_3\in\N+\f{1}{2}$.}
\label{inter}
\end{figure}



\section{3-Dimensional quantum supergravity}
\label{sec:superPR}

\subsection{A topological super Ponzano-Regge model}

As the starting point for the quantization we take the BF-type
formulation of 3-dimensional supergravity (\ref{bfsugra}), without
cosmological constant. In order to perform the quantization we wish to
follow the same recipe as in Section~\ref{sec:qbf}. Superficially,
everything seems to go through (cellular decomposition, parallel
transports etc.), if we simply replace the group $\SU(2)$ by the
supergroup $\OSp(1|2)$. However, there is a problem. An expression
such as (\ref{eq:bfdelta}) or (\ref{eq:bfchar}) is no longer well
defined since we encounter ordering ambiguities between functions on
the supergroup as they no longer commute. On the other hand, the
corresponding circuit diagram representation of the partition function
is still well defined
\cite{Oe:qlgt}. Furthermore, fixing the anomaly one obtains again a
partition function that does not depend on the chosen cellular
decomposition \cite{Oe:qlgt,GiOePe:diagtop}.

Thus, we obtain a good quantization which we write as
\[
 \pfx{sugra}= \hat{\kappa}^p \sum_{V_f} (\prod_f \sdim V_f)\,
 \cdh(V_f) . 
\]
As before, the hats indicate that the quantities involved carry
nontrivial gradings. The sum is now a sum over irreducible
representations $V_f$ of $\OSp(1|2)$ for each face $f$. More
precisely, as explained in Section~\ref{sec:reposp}, we choose only
those representations that obey the spin-statistics relation. This
means, we effectively restrict to the reduced supergroup
$\OSp'(1|2)$ \cite{Oe:qgeosusy} (see Section~\ref{sec:reposp}).
The bosonic theory which is extended is the graded Ponzano-Regge model
(\ref{eq:PRcg}) and not the non-graded one (\ref{eq:PRc}).

In the following we shall consider the appropriate
representations of $\lalg{osp}(1|2)$ as discussed in
Section~\ref{sec:reposp} (see footnote~\ref{fn:osp}).

\subsection{Decomposition into supersymmetric $6j$-symbols}
\label{sec:superdec}

It would be desirable to obtain also in the supersymmetric case a
formulation of the model in terms of the standard spin foam language,
paralleling the usual formulation of the ordinary Ponzano-Regge model
(\ref{eq:PRexpl}). Roughly speaking,
we wish to decompose the cables into
intertwiners (Figure~\ref{fig:njdec}) as in Section~\ref{sec:pr}, to
obtain a sum over products of $6j$-symbols associated with the
tetrahedra of a simplicial decomposition.

Superficially, the procedure seems to be the same as already sketched
in Section~\ref{sec:pr}. There is a crucial difference, however. In
contrast to the $\su(2)$ case the
$3j$-symbols for $\osp(1|2)$ that replace the cable
(Figure~\ref{fig:njdec}) no longer
have a purely combinatorial definition. In other words, the meaning of
a $3j$-symbol now depends on its graphical representation. Another way
to say this is that the $3j$-symbols for $\osp(1|2)$ do not enjoy the
same symmetry properties (under moving around and exchange of legs)
as the ones for $\su(2)$. This
dependence on the graphical representation is inherited by the
$6j$-symbols composed of the $3j$-symbols. One might view this as
causing an information loss in going to a description such as
(\ref{eq:PRexpl}). This phenomenon is well known when employing
quantum groups. Indeed, in the case of the Ponzano-Regge model it is
not sufficient to replace $6j$-symbols with quantum
$6j$-symbols. There is crucial topological information which is not
captured by a formula of the type (\ref{eq:PRexpl}). Although it is
common to write down such a formula also in the quantum group case
it is then implicitly understood that this is not a complete
definition in itself. Unfortunately, this fact does not seem to be universally
appreciated in the physics community.

The same situation (although in a sense less severe) occurs for
supergroups. Thus, although definitions of $6j$-symbols for
$\osp(1|2)$ do exist \cite{susy6j}, it would not be sufficient to
write a naive formula as (\ref{eq:PRexpl}) with those to get a well
defined model.
In contrast to the $\su(2)$ case additional topological information
must be specified. It is a crucial advantage of the circuit diagram
formalism that it does not suffer from this problem but encodes the
complete information \cite{Oe:qlgt}. Thus, when splitting up cables we
need to keep track of this information. In the following we describe
this splitting in more detail.

The procedure consists of a few steps:
\begin{itemize}
\item First, we choose arrows (orientations) for each face of the dual
  2-complex (i.e.\ the spin foam), allowing
to distinguish a representation from its dual.
\item Then we cut each cable in two, decomposing it into pairs of $3j$
  symbols
following Figure~\ref{fig:njdec}. Generically, a cable would decompose
into a sum of such pairs. But in the case of $\osp(1|2)$, as in the
$\su(2)$ case,
there is one unique intertwiner up to normalization between a tensor
product of three irreducible representations.
A cable decomposes simply into two $3j$ symbols.

Each $3j$ symbol corresponds to a map
$R^{j_1}\otimes R^{j_2}\otimes R^{j_3}\arr \C$, or
$R^{j_1}\otimes R^{j_2}\otimes (R^{j_3})^*\arr \C$ etc.\ depending
on the arrows
of the edges around the cable. These invariant tensors can all be constructed
up to normalization from the tensor product decomposition, defining the
maps $R^{j_1}\otimes R^{j_2}\arr R^{j_3}$.

Then, of course, there is the issue of the normalization of these
$3j$ symbols. 
A key property of the cables (which is crucial for the discretization
independence) is that the composition of two cables is equivalent to a single
cable (see Figure~\ref{theta.eps}).
This translates into a normalization of the $3j$ symbols
such that the contraction of the two $3j$ symbols is equal
to 1 (see Figure~\ref{theta.eps}).
Note that all the individual $3j$ symbols are defined with respect to a
fixed graphical representation.

\begin{figure}[t]
\begin{center}
\includegraphics[width=8cm]{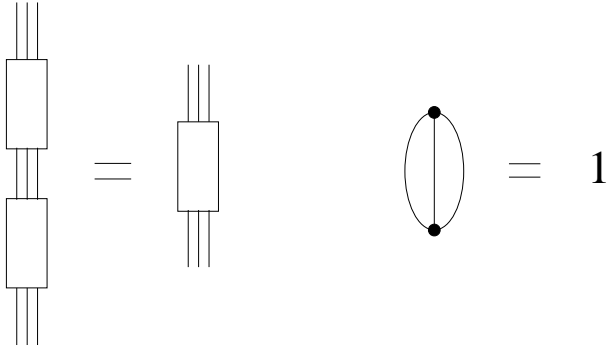}
\end{center}
\caption{The projection property of the cables implies a normalization
condition for $3j$-symbols.
The normalization of the $\Theta$ net to one is valid for all choices
of orientations of the links.}
\label{theta.eps}
\end{figure}

\item After having decomposed the cables associated to the edges, we
  gather and contract the $3j$ symbols around each vertex (i.e.\
  tetrahedron) and evaluate the resulting diagram thus defining a
  $6j$ symbol. Crucially, in this process the rules of
  Section~\ref{sec:circdiag} apply, in particular those of
  Figure~\ref{fig:elemtan}.

  Then the partition function is the product of all these
  $6j$ symbols, 
  together with the measure on the representation defined by the
  superdimension.
\end{itemize}

At the end of the day, we obtain a partition function for a closed manifold
written as
\be
\pfx{sugra}=\sum_{j_f}(\prod_{f} (-1)^{2j_f})\prod_v\{6j\}_{susy}.
\ee
Note that in the process we have explicitly incorporated the
topological matching information into the $6j$-symbols. In particular,
the $6j$-symbols are not necessary defined uniformly. That is,
$6j$-symbols associated with different tetrahedra might have a priori
different normalizations, even if they are labeled with the same
representations. Given the results of \cite{susy6j} it is probable
that this nonuniformity can be removed up to sign factors. While that
would be clearly desirable, it goes beyond the scope of this paper.

Let us point out that the weight for each representation is simply a
sign $(-1)^{2j}$ 
compared to the $\su(2)$ weight of modulus $(2j+1)$. The supersymmetric
partition function is thus likely to be less divergent than the
classical Ponzano-Regge 
partition function (which needs a regularization).


\subsection{Boundaries and TQFT}
\label{sec:boundary}

So far we have only concerned ourselves with the partition
function. For a full-fledged quantum theory we need to specify states,
observables etc. We follow here the approach of topological quantum
field theory (TQFT) which is well established for three-dimensional
quantum gravity \cite{TuVi:inv3,ooguri}. Concretely, we consider
3-manifolds with boundaries. State spaces are associated with these
boundaries while linear maps between them are associated with the
3-manifolds. The construction for quantum supergravity proceeds
essentially identically to the construction for quantum
gravity. Nevertheless, we briefly recall it here to highlight
differences that arise.

Let $M$ be a compact three-dimensional manifold with boundary
$\partial M$. We consider a cellular decomposition of $M$. This
implies in particular a cellular decomposition of $\partial
M$. Consider the dual 1-complex on $\partial M$, consisting of nodes
(dual to 1-cells) and links (dual to 0-cells). This 1-complex bounds
the 2-complex in the interior of $M$ that is dual to the cellular
decomposition of $M$. For a given labeling of faces and edges of the
2-complex in $M$
(in the picture where cables are decomposed as in
Section~\ref{sec:superdec})
we get an induced labeling of links and nodes on the boundary. In
other words, we obtain a spin foam in the interior bounded by a spin
network on the boundary \cite{Bae:spinfoams}. The details of this
construction in
terms of the circuit diagram formalism are described in
\cite{Oe:qlgt}.

A given cellular decomposition $C$ of a 2-manifold $\Sigma$ thus
yields a vector space $H_{\Sigma,C}$, namely the space with basis
given by all labelings of the dual 1-complex with representations and
intertwiners. If $\Sigma$ is the boundary of a 3-manifold $M$ with
cellular decomposition $\hat{C}$, the
partition function $\pfx{sugra}$ gives rise to a linear map
$H_{\Sigma,C}\to \C$. This map is now independent of how $\hat{C}$
extends the cellular decomposition $C$ into the interior of $M$.
Dualizing the vector space $H_{\Sigma,C}$ allows to write the
linear map as $\C\to H^*_{\Sigma,C}$. In particular, 
given a 3-manifold $M$ with boundaries $\Sigma_1, \Sigma_2$
and cellular decomposition yields a linear map $H_{\Sigma_1,
C_1} \to H^*_{\Sigma_2, C_2}$, where $C_1$ and $C_2$ are the induced
cellular decompositions of the boundary components. Note that
orientation reversal of a 2-manifold corresponds to dualization of the
associated vector space (see \cite{Oe:qlgt} for details). Thus a given
2-manifold $\Sigma$ with cellular decomposition $C$ defines a
linear map $P_{\Sigma,C}:H_{\Sigma,C}\to H_{\Sigma,C}$ associated with
the 3-manifold $\Sigma\times I$ where $I$ is an interval. Here $C$ is
extended arbitrarily to the interior of $\Sigma\times I$. Note that
$P_{\Sigma,C}$ is a projector due to discretization independence in
the interior (see \cite{ooguri} for an explicit construction in the
Ponzano-Regge case).
The physical state space associated
with a 2-manifold $\Sigma$ is defined to be
$H_{\Sigma}\defeq P_{\Sigma,C}(H_{\Sigma,C})$. This is independent of
the cellular decomposition $C$. Furthermore,
$P_{\Sigma,C_1}(H_{\Sigma,C_1})$ and $P_{\Sigma,C_2}(H_{\Sigma,C_2})$
are naturally isomorphic by considering the 3-manifold $\Sigma\times
I$ with the two different cellular decompositions $C_1$ and $C_2$ at
the boundaries. The linear map between state spaces for given cellular
decompositions descends to a linear map between the physical state spaces
$\rho_M:H_{\Sigma_1}\to H_{\Sigma_2}$.

So far everything seems to be exactly analogous to the case of
quantum gravity. In particular, the spin networks on the boundary used
to define the state spaces $H_\Sigma$ are as usual graphs with nodes
and links. Representations are associated with links and intertwiners
with nodes. Only now the ``group'' is $\OSp(1|2)$ and not
$\SU(2)$. However, there is a subtlety. Namely, specifying a graph,
representations and intertwiners is not enough to define a concrete
spin network uniquely. There are ordering ambiguities due to the
noncommutativity of the function algebra on $\OSp(1|2)$. Expressing
the labelings (representations and intertwiners) in terms of matrix
element functions it is necessary to specify their ordering. A change
of ordering introduces factors of $-1$ for fermionic components.
Of course this is not a new degree of freedom and does not affect the
abstract definition of the spaces $H_{\Sigma}$. It is rather in
concrete calculations that this ambiguity needs to be taken into account.
In the circuit diagram formalism this problem is coherently dealt
with as the explicit graphical representation takes care of the
ambiguities (see \cite{Oe:qlgt}). As we shall see this issue
resurfaces (as it should) in the discussion of the loop approach.

The standard physical interpretation of the theory is roughly as
follows. Consider two 2-manifolds $\Sigma_1$ and $\Sigma_2$ (thought
of as ``spacelike hypersurfaces'') and a connecting 3-manifold $M$
(thought of as the ``time evolution''). Let $\psi_1$ and $\psi_2$ be
elements of the physical state spaces $H_{\Sigma_1}$ and
$H_{\Sigma_2}$. Then the map
$\rho_M:H_{\Sigma_1}\to H^*_{\Sigma_2}$ can be evaluated on $\psi_1$
and $\psi_2$ yielding a complex number. This gives a ``transition
amplitude'' between the state $\psi_1$ and the state $\psi_2$.
There are certain problems with this interpretation however (notably
to define a measurement process in this context). It has recently been
suggested to consider manifolds with a \emph{connected} boundary
instead and giving a physical interpretation to amplitudes associated
with state spaces on such boundaries
\cite{Oe:catandclock,Oe:boundary}. This implies in particular, that
these boundary surfaces should be thought of as having ``timelike
components'' and that amplitudes are evaluated on a \emph{single}
state. These questions, however, go beyond the framework of the
present paper.


\subsection{Canonical quantization}

Similarly to the pure gravity situation, one can develop a loop quantization
of supergravity, which provides a canonical framework for the super
Ponzano-Regge spin foam model. Indeed, the (kinematical) states of a loop
quantum supergravity will actually be the $\osp(1|2)$ spin networks
which have been identified as boundary states of the super
Ponzano-Regge model. From this point of view, the spin foam model
defines the space-time version, or path integral, for loop quantum
supergravity.
It allows to take into account the dynamics of the canonical theory and
thus to define a projector onto physical states. The canonical framework
is particularly interesting for it introduces geometrical operators
such as a (super)length operator, at the
kinematical level, which allow to probe
the physical meaning of the states of the quantum theory.

We can proceed to the canonical analysis of 3-dimensional
(Euclidean de~Sitter)
supergravity on a three-manifold of the type ${\cal M}=\R\times \Sigma$.
Starting with the super BF action \Ref{bfsugra}, the canonical variables
are the (space components of the) supertriad ${\cal E}$ and the
(space components of the) superconnection ${\cal A}$, both valued in the
superalgebra $\lalg{osp}(1|2)$.
These variables are conjugate and the Hamiltonian consists of two
constraints.
The first one, $d_{{\cal A}}{\cal E}=0$ imposes a vanishing (super)torsion
and generates $\OSp(1|2)$ gauge
transformations. We call it the super-Gauss law.
The second one, ${\cal F}({\cal A})=0$ imposes flatness of the
superconnection.
It generates "topological" gauge transformations of the supertriad\footnote{
The topological transformation reads
$\delta{\cal E}=d_{{\cal A}}\lambda$ for a gauge parameter $\lambda$.}.
It can be decomposed into three pieces, a first one generating
space diffeomorphism (on $\Sigma$), a second one generating
right handed supersymmetry transformations and a last one -the Hamiltonian
constraint- generating the evolution in time.

One can then ``loop quantize'' the theory following the same steps
as for the usual 3-dimensional gravity theory \cite{thiemann:2+1,2+1}.
In this context, one considers the partial observables given by
gauge invariant cylindrical functions of the superconnection ${\cal A}$.
A cylindrical function
depends only on the holonomies of ${\cal A}$ along
the edges of a given closed oriented graph $\Gamma$:
$$
\phi_{\Gamma}({\cal A})=
\phi(\{U_e[{\cal A}],e\in\Gamma\}).
$$
Then one would like to impose $\OSp(1|2)$ gauge invariance, i.e.\ invariance
under $\OSp(1|2)$ at the vertices $v$ of the graph $\Gamma$. Heuristically,
following the procedures developed in the pure gravity case,
this would read as:
$$
\forall k_v\in \OSp(1|2),\,
\phi(\{k_{s(e)}^{-1}U_e[{\cal A}]k_{t(e)},e\in\Gamma \})=
\phi(\{U_e[{\cal A}],e\in\Gamma\}),
$$
where $s(e)$ and $t(e)$ respectively denote the source and target vertices
of an edge $e\in\Gamma$.
In fact, a function of the (super)holonomies $U_e$ is actually defined as
a function of the matrix elements $t(U_e)$ of the (super)group elements.
Then, one must be careful since these matrix elements do not commute:
we must choose a (full) ordering of all the edges $e$. And
the precise definition
of gauge invariance will depend on the chosen convention.
This is the same problem that one faces when dealing with quantum groups,
except the ordering ambiguities only introduce signs in our case,
which makes the whole situation (much) simpler.

Using the Haar measure $d\mu$ on $\OSp(1|2)$ \cite{haar}, one can introduce
the measure product of $d\mu$ on all edges $e\in\Gamma$, which
defines a natural scalar product on the space of square integrable
cylindrical functions.
A basis of the resulting Hilbert space is provided by the
$\osp(1|2)$ {\it super spin networks} \cite{lqsg,spinnet}.
These are labelled
by a $\osp(1|2)$ representation for each edge and
a $\osp(1|2)$ intertwiner
for each vertex: $|\Gamma,j_e,i_v\ra$.
Let us point that these super spin networks actually depend on the ordering
chosen for the edges of the graph. Nevertheless, as the ordering defines
the gauge invariance condition, it does not introduce further states in
our basis.
Let us have a look at an example.
We choose a particular ordering of the edges,
then a spin network functional will read:
$$
\phi(U_1,..,U_E)=t(U_1)t(U_2)..t(U_E)\times\textrm{ intertwiners}.
$$
As $\OSp(1|2)$ gauge invariance is imposed on the intertwiners, if one
would like to check how it reads on the edge 1, then one needs to translate
the conditions from the intertwiners to the matrix element $t(U_1)$ commuting
it with all the other $t(U_e)$, which introduces signs each time the matrix
elements have an odd parity.

Piling up all these Hilbert spaces for each graph $\Gamma$
(i.e summing over graphs $\Gamma$), 
one constructs a Hilbert space of $\OSp(1|2)$ invariant states of the
superconnection,
which can be considered as
the Hilbert space of square integrable functions over gauge equivalence
classes of (generalized) superconnections.

One last remark on $\lalg{osp}(1|2)$ spin networks is to point out that, as
all $R^j$ representations are symmetric/antisymmetric tensor products
of the fundamental $R^{1/2}$ representation, one can
decompose any super spin network into a superposition of fundamental
loops -labelled by $R^{1/2}$- by splitting each $R^j$ labelled edge
into a symmetrized/antisymmetrized system of ropes, each rope  corresponding
to a $R^{1/2}$ representation \cite{spinnet}.

To finish the quantization, one needs to impose the flatness constraint
${\cal F}({\cal A)}=0$. First, one gets diffeomorphism invariant states
by considering equivalence classes of graphs under diffeomorphisms
of $\Sigma$. Then, one would like to implement the (right handed) supersymmetry
constraint and the Hamiltonian constraint on the resulting Hilbert space.
In the present work, we are not going to discuss this issue directly within
the canonical framework.
Instead, we tackle this issue from the {\it spin foam} point of view:
we construct the path integral for the theory, which, once interpreted
within the canonical framework, provides us with a projector onto
the physical Hilbert space (of states satisfying all the constraints).
Indeed, the super Ponzano-Regge model, implementing  the path
integral of the super BF action \Ref{bfsugra},
is a topological state sum
whose boundary states (kinematical states) are actually the $\lalg{osp}(1|2)$
spin networks and which defines a projector
on the physical states of supergravity (see Section~\ref{sec:boundary}).

An attractive aspect of the Hilbert space of (super) spin networks is
its geometrical interpretation. Indeed, adapting techniques of
$2+1$-dimensional loop quantum
gravity \cite{2+1} just like it is possible to extend results
of ($3+1$) loop quantum gravity to loop quantum supergravity \cite{lqsg},
super spin networks appear as eigenvectors of the {\it super length} operator.
More precisely, the super length of a curve
$c:\tau\in [0,1]\rightarrow c(\tau)\in\Sigma$ is:
\be
L_c=\int_{[0,1]} d\tau\ \sqrt{
\dot{c}^a(\tau) \dot{c}^b(\tau)\,
STr_{osp(1|2)}({\cal E}_a(c(\tau))\, {\cal E}_b(c(\tau)))
}.
\ee
One can quantize it as an operator $\what{L_c}$ by replacing
${\cal E}$ by the derivative with respect to ${\cal A}$ (and
regularizing the expression) just as the gravity case.
Then its action is diagonal on super spin networks. The eigenvalue
depends on the $\lalg{osp}(1|2)$
representations living  on the edges of the spin
network intersected by $c$ and is given by the sum of the squareroot
of the Casimir of these representations:
\be
\what{L_c}|\Gamma,j_e,i_v\ra=
\left(\sum_{e|e\cap c\ne\emptyset}
\sqrt{j_e\left(j_e+\f{1}{2}\right)}
\right)
|\Gamma,j_e,i_v\ra.
\ee
One should be careful that this super length operators depends on both the
metric field and the fermionic fields.
It is not the simple length operator $\what{l_c}$:
\be
l_c=\int_{[0,1]} d\tau\ \sqrt{
\dot{c}^a(\tau) \dot{c}^b(\tau)\,
Tr_{su(2)}(e_a(c(\tau))\, e_b(c(\tau)))
}.
\ee
Indeed, $l_c$ is invariant under $\SU(2)$ but not under $\OSp(1|2)$, so that
its action cannot be implemented on a space of $\OSp(1|2)$
invariant functionals. Considering a small curve intersecting
a single edge $e$ labelled with the representations $R^{j}$, $\what{l_c}$
would be the square root of the $\SU(2)$ Casimir operator $J^iJ^i$ on $R^{j}$
and would distinguish the two isospins $V^j$ and $V^{j-1/2}$ assigning
them different length values $\sqrt{j(j+1)}$ and
$\sqrt{(j-\f{1}{2})(j+\f{1}{2})}$. Then the fermionic fields
contribute $-\f{j}{2}$ (resp. $\f{1}{2}(j+\f{1}{2})$)
to the Casimir on  $V^j$ (resp. $V^{j-1/2}$), so that the super length
is diagonal.

From this perspective, one can view
super spin networks as supersymmetric superposition of (pure) geometry
states given by the usual $\su(2)$ spin networks (labelled with $\su(2)$
representations $V^j$).


\subsection{Identifying Fermions on the spin foam}

As the Ponzano-Regge model provides us with a path integral formulation for
three-dimensional quantum gravity, the super Ponzano-Regge model based on
$\osp(1|2)$ provides us with a path integral for three-dimensional
quantum supergravity.
In this context, it would be interesting to distinguish the gravity
degrees of freedom from the fermionic degrees of freedom in our
supersymmetric state sum model. In particular, we know that fermions
live at the nodes of spin networks in standard loop quantum gravity
\cite{lqgmatter}. However, in the present super Ponzano-Regge model the
fermionic field is taken into account in the holonomies of the super
connection ${\cal A}$ and therefore in the representations (of
$\osp(1|2)$) living on the links of the super spin networks. It would
be enlightening to reformulate the super Ponzano-Regge model
with "fermions" at the nodes in such a way as to make the link with
the usual loop quantum gravity formalism.

In the following, we propose such a picture in which we can
identify objects similar to fermion worldlines on the spin foam.
It looks like some fermions propagating on some gravity background and
therefore the super Ponzano-Regge model appears as a path integral for
gravity plus fermions. Nevertheless, we do not prove
that the proposed picture is "right", in the sense that the paths drawn
on the spin foam truly correspond to the fermionic degrees of
freedom. To verify this conjecture one would have to
to study a continuum limit (through
asymptotics of the supersymmetric $6j$-symbols), or a
semi-classical limit, and we postpone this analysis to future work.

Let us start by summarizing the geometrical structure of the (super)
Ponzano-Regge model. It is usually formulated on a simplicial
triangulation of a three-dimensional manifold: we work with tetrahedra
glued along common triangles. We can extend this to a generic cellular
decomposition made of 0-cells (points), 1-cells (lines), 2-cells and
3-cells. In the spin foam context, we consider the dual structures:
$$
\begin{array}{lcl}
 {\rm \bf spin foam} & \leftrightarrow &{\rm \bf triangulation} \\
{\rm vertices} &\leftrightarrow& {\rm 3-cells} \\
{\rm edges} &\leftrightarrow& {\rm 2-cells} \\
{\rm faces} &\leftrightarrow& {\rm 1-cells} \\
{\rm volumes} &\leftrightarrow& {\rm 0-cells} 
\end{array}
$$
so that vertices are linked by edges corresponding to the 3-cells
glued along 2-cells.
Usually,  we consider only the dual 2-complex, meaning  that we forget
about the last line of the above correspondence, "losing" the notion
of dual volumes (and therefore of points in the triangulation).

Now, in the state sum model, we attach (super) group representations
to the 1-cells of the triangulation and intertwiners to the 2-cells
(intertwining the representations living on its boundary edges). In the spin
foam setting, representations are attached to faces and intertwiners
live on the edges (intertwining the representations of the faces
incident on a given edge). This way, taking a generic slice of the
spin foam, we obtain a graph with representations on its links (cut
faces) and intertwiners at its nodes (cut edges), i.e.\ a spin network.

\begin{figure}[t]
\begin{center}
\includegraphics[width=5cm]{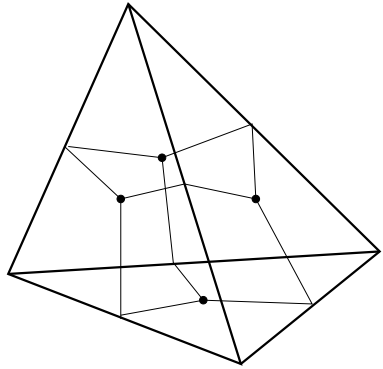}
\end{center}
\caption{Tetrahedron and its boundary spin network.}
\label{tetraboundary}
\end{figure}

Now the amplitude of a spin foam configuration is the product of
amplitudes assigned to each vertex (3-cell) multiplied with
weights associated to the edges (2-cells) and faces
(1-cells). The weight of a face is the (super) dimension of the
representation.
The amplitude of a vertex is given by the evaluation of its \emph{
boundary spin network}.
More precisely, we consider the 3-cell around a vertex. Its
boundary is a sphere (consisting of several 2-cells).
The intersections of the sphere with the faces around
the vertex form a (super) spin network (on the sphere). This boundary
spin network can also be seen as the 1-skeleton of the dual complex of the
boundary of the 3-cell.
In the case of a simplicial decomposition, the 3-cells are the tetrahedra
and the boundary spin networks can be identified with the $6j$-symbols
(see Figure~\ref{tetraboundary}).

In the following, we will restrict our analysis to 3-valent spin
networks, i.e.\ the 2-cells of the triangulation are all triangles. As it
is straightforward to decompose an $n$-valent intertwiner into
3-valent intertwiners (the same way that a generic polygon can be decomposed
into internal triangles), the whole construction will be easily
extended to the generic case.

Let us forget about sign ambiguities for the moment and analyze the
structure of $\osp(1|2)$ spin networks. Comparing to the Ponzano-Regge
model, we expect pure gravity degrees of freedom to be encoded in
$\su(2)$ structures. It is then natural to decompose each $\osp(1|2)$
representation into its $\su(2)$ isospins. Applying this decomposition
to an $\osp(1|2)$ spin network, we must distinguish two types of
$\osp(1|2)$ intertwiners:
\renewcommand{\theenumi}{\roman{enumi}}
\renewcommand{\labelenumi}{(\theenumi)}
\begin{enumerate}
\item Intertwiners intertwining representations $R^{j_1}$, $R^{j_2}$,
$R^{j_3}$ with $j_1+j_2+j_3\in\N$, in which case there exists a
component intertwining between the isospins $V^{j_1}$, $V^{j_2}$,
$V^{j_3}$; \label{boso}
\item Intertwiners intertwining representations $R^{j_1}$, $R^{j_2}$,
$R^{j_3}$ with $j_1+j_2+j_3\in\N+\f{1}{2}$, in which case there doesn't exist
any $\su(2)$ intertwiner between $V^{j_1}$, $V^{j_2}$ and $V^{j_3}$.
\label{ferm}
\end{enumerate}
It appears natural to call an intertwiner of the type 
\Ref{boso} a bosonic intertwiner, or {\it boson}, and an intertwiner
of the type \Ref{ferm} a fermionic intertwiner, or {\it fermion}.

To analyze the structure of an $\osp(1|2)$ spin network, let us
introduce a graphical convention for the $\su(2)$
spin networks resulting from the decomposition of the $\osp(1|2)$
spin network:
solid and dotted lines corresponding respectively to upper and lower
isospins, $V^j$ or  $V^{j-\f{1}{2}}$, of the original $R^j$ representation.

\begin{prop}
Considering a 3-valent $\osp(1|2)$ spin network, it contains an even
number of fermionic intertwiners (fermions) and of bosonic
intertwiners (bosons).
Decomposing each $\osp(1|2)$ representation into its two $\su(2)$
isospins, an upper one and a lower one,
the $\osp(1|2)$ spin network appears as a sum of $\su(2)$ spin
networks, which we draw with solid and dotted links corresponding to
upper and lower $\su(2)$ representations. Then, in each
of the resulting $\su(2)$ spin networks, dotted links form lines on the
spin network linking the fermions: the set of dotted lines can be
partitioned into lines between pairs of fermions (each fermion being linked
to a single other fermion) and closed loops.
\end{prop}

Let us first point out that a 3-valent graph has an even number of
nodes\footnote{If $E$ is the number of links of the graph and $V$
its number of nodes then $2E=3V$.}.
The key point of the
proof is that there is an even number (0 or 2) of dotted lines at a
bosonic intertwiner and an odd number (1 or 3) at a fermionic
intertwiner (as drawn in Figure~\ref{inter}).
Then, from the point of view of the dotted lines, we can forget the solid 
lines and the bosonic intertwiners, and only consider the graph made of 
the dotted links and the fermionic intertwiners, which we call the dotted 
graph.

Among these fermionic 
intertwiners, there are some which are attached to only 
one dotted line. Such a fermion $f_1$ is linked directly to another 
fermion $f_2$
through that dotted line. Thus, we can put aside that couple of fermions,
$f_1$ and $f_2$, 
and that dotted line. The dotted graph left after removing these two 
fermions is still a graph of the same type, since $f_2$ had either one 
attached dotted link (the one linking it to $f_1$) or three dotted links in 
which case $f_2$ becomes a bivalent node on the dotted graph and we can 
forget it. After having removed all such fermions, either there are none 
left and the dotted graph is a set of loops,
or we are left with a 3-valent graph, which we call the reduced dotted 
graph.

Then any 3-valent graph has an even number of nodes and it is moreover
possible to show by induction that we can decompose it into lines
connecting pairs of vertices and (closed) loops.

\begin{figure}[t]
\begin{center}
\psfrag{b}{$b$}
\psfrag{f}{$f$}
\psfrag{e}{=}
\psfrag{p}{+}
\psfrag{ddd}{\dots}
\includegraphics[width=12cm]{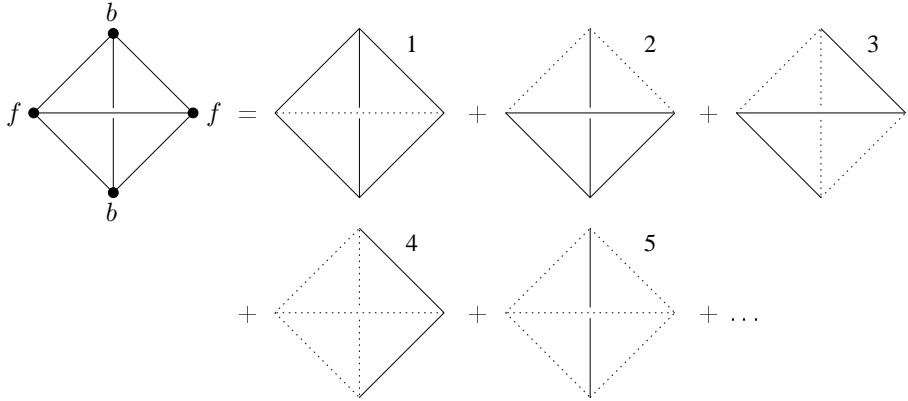}
\end{center}
\caption{Decomposition of the tetrahedral net with two fermionic
  intertwiners on $\su(2)$ spin networks with solid/dotted lines for
upper/lower $\su(2)$ isospins (the dots mean summing over the
symmetric situations). One sees that the dotted links form
a path between the two fermions, plus possibly a loop.  Also, one
notices, in the case of the diagram~5 for example,
that the choice path/loop is not unique and that one can even forget the
loop and consider a single long dotted line linking the two fermions.}
\label{tetra2f}
\end{figure}

\begin{figure}[t]
\begin{center}
\psfrag{f}{$f$}
\psfrag{e}{=}
\psfrag{p}{+}
\psfrag{ddd}{\dots}
\includegraphics[width=12cm]{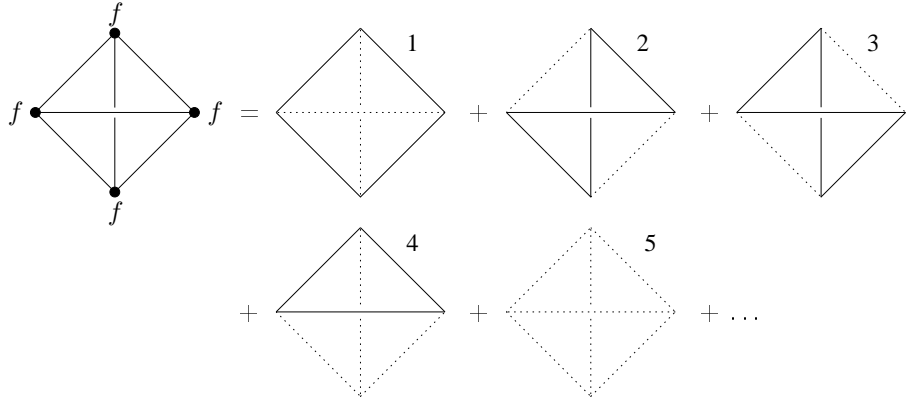}
\end{center}
\caption{Decomposition of the tetrahedral net with four fermions: The
  dotted links form two lines, each one linking two fermions together.
In the diagrams 1,2,3, it is straightforward to see which fermion is
linked with which other. In the diagrams 4,5, the situation is more
complicated and it is just a question of interpretation to decide
which fermionic intertwiner is linked with which other one, i.e.\
which fermion goes where.}
\label{tetra4f}
\end{figure}

Let us point out that the decomposition of the dotted lines -linking
pairs of fermions- and loops is not unique at all.
 Therefore, there is a choice of "interpretation" in the choice of 
which fermion is linked to which other one. We illustrate these 
considerations 
with the example of the tetrahedron net ($6j$-symbol) in
Figure~\ref{tetra2f} and Figure~\ref{tetra4f}.


Now, let us come back to the spin foam model.
An $\osp(1|2)$ spin foam is labelled by $\osp(1|2)$ representations
on its faces, and the intertwiners live on its edges, which are then
either bosonic or fermionic.
The amplitude of a given
$\osp(1|2)$ spin foam is the product of the boundary spin networks
corresponding to its 3-cells/dual vertices. Each of these boundary
spin networks is the sum of many $\su(2)$ spin networks on which are
drawn loops and lines linking each fermionic intertwiner to another
one.
Let us emphasize that each edge corresponds to a cable that we cut into
two $\osp(1|2)$ intertwiners which are the same (up to some sign conventions),
so that both are fermionic or both are bosonic. This property is inherited by
the
$\su(2)$ intertwiners, drawn with solid and dotted lines, when decomposing
the $\osp(1|2)$ representations into $\su(2)$ isospins.

The resulting space-time picture is that we have fermions propagating
on the edges of the spin foam and meeting at the vertices. Each
term of the sum describes what happens at the vertices (fermion
scattering,\dots) and therefore can be understood as a {\it Feynman
  diagram} for fermions drawn on the geometrical skeleton of the
spin foam.
At the level of the amplitude, the full amplitude is the sum of
the amplitudes of $\su(2)$ spin foams on which are drawn fermionic loops
and lines.

On the triangulation, intertwiners live on the triangles, which are
thus fermionic or bosonic. Then, each term of the boundary spin
network of a given 3-cell tells one which fermion (fermionic triangle)
goes where. And one can build a sequence of triangles by following a
given fermionic triangle through every 3-cell,  simply the
path of a fermion.

Through this construction, we have identified fermionic degrees of
freedom attached to the intertwiners, which matches with the canonical
loop quantum gravity picture. Moreover, we have interpreted the super
Ponzano-Regge model as providing a path integral for gravity+fermions,
in which the space-time is described as fermions propagating on a
"background" geometry described with the usual $\su(2)$ representations
of the Ponzano-Regge model. Let us point out that it is not simply a
background geometry since the fermions modify the geometrical
properties -the length of curves for example- along their path.

To sum up, given a triangulated three-dimensional manifold $\Delta$,
one labels its
1-cells with $\osp(1|2)$ representations and assigns an amplitude
to each configuration (consistent labelling under the
recoupling theory of $\osp(1|2)$). Then the full super Ponzano-Regge
partition function for $\Delta$ is the sum of these amplitudes over
all configurations. Considering a given term (fixed $\osp(1|2)$ labels),
decomposing each $\osp(1|2)$ representations into $\su(2)$
representations, the amplitude appears like a (supersymmetric)
superposition of pure gravity/geometry space-time states/configurations
defined by $\su(2)$ edge labels like in the Ponzano-Regge model. This
superposition of gravity states can also be interpreted as one pure gravity
state plus gravity states with propagating fermions.

This can be written as a decompositions of the partition function
for a fixed closed triangulated manifold $\Delta$:
\be
Z(\Delta)=\sum_{J_{\osp(1|2)}}Z_{(SPR)}(\Delta,J)
=\sum_{J_{\osp(1|2)}}\sum_{j_{\su(2)}\harr J} c(J,j)Z_{(\su(2))}(\Delta,j).
\ee
$Z_{(SPR)}$ is the super Ponzano-Regge
amplitude for specified colorings $J_{\osp(1|2)}$.
When we decompose $\osp(1|2)$ representations into $\su(2)$ isospins, this
amplitude reads as a sum of $\su(2)$ terms, with
$c(J,j)$ some normalization coefficients
(such that the sum is supersymmetric).
Each term $Z_{(\su(2))}(\Delta,j)$ is a product of 6j-symbols
of $\su(2)$
and the coefficients $c(J,j)$ can be considered as part of
the measure for the sum. Then, using our graphical notation
with solid/dotted lines, we can identify the fermions, living on triangles,
and the fermionic lines
corresponding to each term $c(J,j)Z^{(PR)}(\Delta,j)$ and it turns out
that a term is non-vanishing iff the fermionic lines form loops
on the spin foam. Therefore, one could swap the sums
on $\osp(1|2)$ and $\su(2)$ labels in the partition
function. This means first summing  on $\su(2)$ representation labels
and then on the possible
consistent ways of embedding them
into $\osp(1|2)$ representations, i.e. over consistent choices of
solid/dotted lines in the evaluation of the 3-cell boundary spin networks.
These consistent configurations will then
be defined by all possible sets of (closed) loops, the fermionic paths,
drawn on the spin foam. Let us emphasize that the sum over $\su(2)$ is
a generalised one: we do not sum over one representation label $j$ for
each face of the spin foam (like in the Ponzano-Regge model) but we
sum over couples of labels -$j$ and $j-\f{1}{2}$- for each face and
allow different labels -$j$ or $j-\f{1}{2}$- around a single face, one
for each adjacent vertex (equivalently, at the level of the
triangulation, a representation $j$ is attached to a 1-cell and we
sum over assignements of $j$ or $j-\f{1}{2}$ to each 3-cell adjacent
to that 1-cell).
Then, when $\Delta$ has boundaries, one must first check which boundary nodes
are fermionic and then the ``loops'' must also include lines drawn between
pairs of boundary fermions.
At the end of the day, the partition function looks like a sum over
gravity configurations and fermion paths drawn on the spin foam, or
equivalently gravity+fermions Feynman diagrams.

Finally, we emphasize that the whole construction is just a "nice"
picture for the moment and that we have not provided a proof that the
fermionic intertwiners are truly physical fermions and that the dotted lines
are truly fermion paths.
It is true that this interpretation seems natural from the point of
view of the representation theory and that it leads to an interesting
understanding of the partition function as a sum over coupled gravity+fermions
configurations.
Nevertheless, only a semi-classical analysis or the study of a
continuum limit could settle the issue and prove the picture right or
wrong.


\subsection{Including a cosmological constant:
$U_q(\osp(1|2))$ spin foams}
\label{sec:ospq}

In spin foam models for quantum gravity, quantum deforming the gauge group
has two uses. On the one hand, it regularizes the state sum by
making finite the sum over representations.
On the other hand, it is supposed to take into account a positive
cosmological constant.
In the case of the Ponzano-Regge model, quantum deforming
$\SU(2)$ into $\SU_q(2)$ leads to the Turaev-Viro model \cite{TuVi:inv3}.
Choosing $q= e^{\f{2\pi i}{N}}$ to be a root of unity
reduces the number of inequivalent
irreducible representations to be finite with a spin cutoff
$j\le (N-2)/2$: Having finite sums makes the Turaev-Viro model
well-defined. Let us recall the q-dimensions of the representations,
which enter the state sum instead of the ordinary dimensions:
\be
\qdim(j)=\f{q^{j+\f{1}{2}}-q^{-j-\f{1}{2}}}{q^{\f{1}{2}}-q^{-\f{1}{2}}}
=\f{\sin\left((2j+1)\f{\pi}{N}\right)}{\sin\left(\f{\pi}{N}\right)}
\xrightarrow[q\arr 1]{} 2j+1 .
\ee
Moreover, it has been shown that the cosmological constant
$\Lambda$ is linked with the parameter $q$ through
$N\sqrt{\Lambda}=2\pi$.
Mathematically, this comes from the reformulation of three-dimensional
quantum gravity
as the sum of two $\SU(2)$ Chern-Simons theories.
More precisely,  the partition function of the Turaev-Viro model,
as the quantization of BF theory with cosmological constant, appears
to be the squared absolute value of the Chern-Simons amplitude
(with level $N$) \cite{roberts}.
On a more empirical level, it can be seen from the asymptotics of
the quantum $6j$-symbols which turns out to be  
the cosine of the Regge action
for (discretized) gravity with cosmological constant.

In this context, it is very appealing to q-deform the super Ponzano-Regge
model writing a ``super Turaev-Viro'' model based on the quantum group
$U_q(\osp(1|2))$.
This is a very simple quantum (super)group.
Just like for all the $U_q(\osp(1|2n))$, there exists an invariant Haar
measure and a Peter-Weyl decomposition \cite{ospq,ospqn}. Tensor
products
of irreducible representations are totally reducible, so that one can
straightforwardly write a corresponding topological model using the usual
techniques to deal with quantum groups. In particular, the circuit
diagram formalism generalizes to this setting \cite{Oe:qlgt}.

Then, for $q= e^{\f{2\pi i}{N}}$ an
$N$th root of the unity, there is once again a finite number
of irreducible representations $R^{j,\pm}$ labelled
by half-integers $0\le j\le (N-1)/2$ and a parity $\pm$.
Choosing, as we did for $\osp(1|2)$, the parity to be the physical one
(i.e.\ to obey the 
spin-statistics relation) their q-superdimension is:
\be
\qsdim(R^j)=(-1)^{2j}\f{q^{2j+1}+q^{-2j}}{q+1}
=(-1)^{2j}\f{\cos\left((4j+1)\f{\pi}{N}\right)}
{\cos\left(\f{\pi}{N}\right)}
\xrightarrow[q\arr 1]{} (-1)^{2j}.
\ee
Therefore, in this setting, sums over representations are finite and
the resulting state sum is well-defined. For more details on the
representation theory and more particularly on the tensor products of
representations, we refer to \cite{ospq,ospqn}.

Paralleling the situation for gravity,
three-dimensional (Euclidean de~Sitter)
supergravity can be formulated
as the sum of two $\OSp(1|2)$ Chern-Simons theories, as explained in
Section~\ref{csugra}.
Thus, its
quantization should yield the quantum group $U_q(\osp(1|2))$ with $q$
related to the cosmological constant as above. Furthermore, the relation
between quantized Chern-Simons theory and the Turaev-Viro model is the
same for super (quantum) groups as for ordinary (quantum)
groups. Thus, the super Turaev-Viro model should indeed encode the
quantization of the supersymmetric theory with cosmological constant.
It would be very interesting to work out the asymptotics of the
super quantum $6j$-symbols to check the relation
$q\leftrightarrow \Lambda$ on a semi-classical level.

Let us point out a weakness in the link between the
q-deformation and the cosmological constant:
the $q$-deformation does not appear naturally in the loop quantum gravity
setting. Indeed, the ``kinematical states'' of quantum gravity
are $\SU(2)$ spin networks and
the cosmological constant $\Lambda$ appears only in the Hamiltonian constraint
governing the dynamics. The same occurs for supergravity.
Through the q-deformation of the Ponzano-Regge model into Turaev-Viro model,
it seems that, at the spin foam level, the $q$-deformation allows to take
into account properly the dynamics implied by the cosmological constant.
We expect the same situation in supergravity.
Still, it would be very interesting to study how the $q\leftrightarrow\Lambda$
relation could appear in the canonical loop gravity framework, both in
the gravity case and in the supergravity case.


\subsection{Towards a Lorentzian model: $\osp_L(1|2)$ spin foams}

Up to now, we have dealt with the Euclidean Ponzano-Regge model
corresponding to a quantization of Euclidean (or Riemannian) gravity.
We can also study the Lorentzian version \cite{davids,laurent} quantizing
$2+1$-dimensional gravity and extend to a supersymmetric Lorentzian
Ponzano-Regge model
which would be a state sum for Anti-de~Sitter supergravity in $2+1$
dimensions.

Let us start by describing the Lorentzian Ponzano-Regge model. As the
gauge group
(in the connection formalism) of $2+1$-dimensional gravity is
$\SO(2,1)$ instead of $\SO(3)$,
we build a model invariant under $\SU(1,1)$ instead of
under $\SU(2)$.
It uses the unitary irreducible representations of $\su(1,1)$, which
are infinite dimensional.
These representations are of two kinds (restricting ourselves to the
principal representations,
appearing in the Plancherel formula for $\su(1,1)$): a continuous
series and two
discrete series, one negative and one positive.
The representation spaces are similar to the $\su(2)$ case, with a
basis $|m\ra$ labelled
by half-integers. The representations of the continuous series
have a basis labelled by all $m\in \Z$ (or $m\in \Z+1/2$ depending on
their parity) whereas
the representations of the discrete series are in a way
complementary to the $\su(2)$
representations and have a basis spanned
by a bounded label, $m\ge j$
(for the positive representation $V^j_{(\su(1,1))}$) or by
$m\le -j$ (for the negative series).
The difference between $\su(2)$ and $\su(1,1)$ can be understood on
the Lie algebra level as arising from a difference in
the hermiticity conditions: For $J_\pm$ defined in \Ref{commrelations},
$\su(2)$ is characterized
by $J_\pm^\dagger=J_\mp$ whereas $\su(1,1)$ is defined by
$J_\pm^\dagger=-J_\mp$ (for more details, see for example \cite{2+1}).

Then, these $\su(1,1)$ representations label the edges of
the triangulation
(or the faces of the spin foam) and have a clear geometric interpretation:
the representations of the continuous series correspond to
space-like edges and
the representations
of the positive (resp. negative) discrete series to future
(resp. past) oriented time-like
edges. Following \cite{laurent},
one can then write a topological model based on the whole category of
continuous
and discrete representations.
However, one can also build a "restricted" topological model using
solely the positive discrete series
(because this set of representations is closed under tensor product).
This ``restricted'' model is particularly interesting for
it contains only oriented time-like representations
(corresponding to time-like edges)
and is interpreted using time-like boundaries\footnotemark.

\footnotetext{In the framework of the restricted model,
there is a natural light cone structure on the spin foam.
Indeed, considering a point in the triangulated manifold,
the set of its future events is given by all points linked to it by a sequence
of future oriented edges.
In this context,
one can easily identify the double cones (or diamonds) and study the
corresponding observables. This brings the model close to the framework of
algebraic quantum field theory.}

Now, we would like to build a supersymmetric Lorentzian Ponzano-Regge model.
Following section \ref{sec:sugra},
it would correspond classically to Anti-de~Sitter (Lorentzian) gravity,
using the gauge group $\OSp_L(1|2)$, the
supersymmetric extension of $\SU(1,1)$, instead of $\OSp_E(1|2)$.
We would only extend supersymmetrically the (positive) discrete series of 
representations and give
a supersymmetric version of the restricted Lorentzian model. Its
natural interpretation would be
with time-like boundaries and fermions on these time-like boundaries. 

Mimicking the representation theory of $\osp_E(1|2)$
(choosing weights $m\ge j$ instead of $|m|\le j$ and rotating by $i$
the generators
$J\pm$), we find representations
$R^j$ of $\osp_L(1|2)$ which are sums of two (unitary, infinite
dimensional) representations,
$V^j$ and $V^{j-\f{1}{2}}$, of $\su(1,1)$.
Each $V^k$ representation can be written in a basis similar to the 
$\su(2)$ case $|k,m\ra$, but with weights $m\ge k$.
The explicit action of the generators is:
\bes
J_3|j,j,m\ra&=&m|j,j,m\ra,\nonumber \\
J_3|j,j-\f{1}{2},m\ra&=&m|j,j-\f{1}{2},m\ra,\nonumber \\
J_\pm|j,j,m\ra&=&
\left(m(m\pm1)-j(j+1)\right)^{1/2}|j,j,m\pm1\ra,\nonumber\\
J_\pm|j,j-\f{1}{2},m\ra&=&
\left(m(m\pm1)-(j-\f{1}{2})(j+\f{1}{2})\right)^{1/2}
|j,j-\f{1}{2},m\pm1\ra,\nonumber\\
Q_+|j,j,m\ra&=&-(m- j)^{1/2}
|j,j-\f{1}{2},m+\f{1}{2}\ra,\nonumber\\
Q_+|j,j-\f{1}{2},m\ra&=&\f{i}{2}(m+j+\f{1}{2})^{1/2}
|j,j,m\pm\f{1}{2}\ra,\nonumber\\
Q_-|j,j,m\ra&=&-i(m+j)^{1/2}
|j,j-\f{1}{2},m-\f{1}{2}\ra,\nonumber\\
Q_-|j,j-\f{1}{2},m\ra&=&-\f{1}{2}
\left(m-\left(j+\f{1}{2}\right)\right)^{1/2}|j,j,m-\f{1}{2}\ra.
\ees
The commutation relations are:
\bes
[J_3,J_\pm]=\pm J_\pm  \qquad [J_+,J_-]=-2J_3 \nonumber\\
\left[J_3,Q_\pm\right]=\pm\f{1}{2}Q_\pm \qquad [J_\pm,Q_\pm]=0 
\qquad [J_\pm,Q_\mp]= -iQ_\pm  \nonumber \\
\{Q_\pm,Q_\pm\}=\mp\f{i}{2}J_\pm \qquad   \{Q_\mp,Q_\pm\}=\f{1}{2}J_3.
\ees
The ``hermiticity'' relations are $J_3^\dagger=J_3$, $J_\pm^\dagger=J_\mp$
and $Q_+^\dagger=\pm Q_-$, $Q_-^\dagger=\mp Q_+$, the sign reflecting
the parity
assigned to the representation just as in the $\osp_E(1|2)$ case.

The (3-valent) intertwiners can be defined through the
decomposition of the tensor product of two representations.
In the $\su(1,1)$ case, it reads as:
\be
V^j\otimes V^k=\bigoplus_{l\ge j+k} V^l,
\ee
where $l$ goes by integer steps. The condition $l\ge j+k$ is the anti-triangular
inequalities for a triangle with only timelike edges.
We get the same structures for the supersymmetric extension:
\be
R^j\otimes R^k=\bigoplus_{l\ge j+k} R^l,
\ee
where $l$ now goes through all half-integer values.

We are not going to describe the details of explicitly constructing the
Lorentzian
super-Ponzano-Regge model. Just as for the Lorentzian Ponzano-Regge model,
we would need to tackle the difficulties linked with the use of a
non-compact group and infinite dimensional representations. This is
beyond the scope of the present paper. Nevertheless, we already see by the
similarity of the representation theories that all
the considerations made for the Euclidean case will translate to the Lorentzian
case: we can also distinguish fermionic and bosonic intertwiners in the
Lorentzian model and a Lorentzian
super-Ponzano-Regge state sum will have a natural interpretation
as a spin foam model with fermions on timelike boundaries.



\section{Conclusions and Outlook}

We have presented a general diagrammatic formalism to deal with
supersymmetric spin foam models. It is on the one hand a special case
of the formalism presented in \cite{Oe:qlgt} and on the other hand an
extension to the graded case of the connection formulation of spin
foam models. Using the formulation of three-dimensional supergravity
as a topological BF theory with an $\OSp(1|2)$ gauge group, we have defined
a super-Ponzano-Regge model, a discretized path integral
for quantum supergravity. The model, extension of the original
Ponzano-Regge model, is naturally independent of the discretization.
We have described the extension of the model to a topological quantum
field theory, thus providing a description of all the elemets of the
quantum theory of supergravity in dimension three.
We have also studied aspects of canonical loop quantum supergravity,
expressed its kinematical states as super spin networks and considered
its relation to our super-Ponzano-Regge model.

Since supergravity contains fermionic degrees of freedom, we have
considered the problem of identifying these in the super-Ponzano-Regge
model. It turned out
that $\osp(1|2)$ intertwiners are of two kinds, which we tentatively
called ``bosonic'' and ``fermionic''. Starting with this assumption,
we can identify (closed) fermion paths on the spin foam.
We can also write
the partition function of the supersymmetric model as
a sum over (supersymmetric) superpositions of pure gravity/geometry states,
which we suggest to interpret as a sum over gravity
configurations plus fermion paths drawn onto some background geometry (defined
by a pure gravity state).

We would need to study in more details how to extend the present procedures
to the Lorentzian case, therefore studying (Anti-de~Sitter) Lorentzian
quantum supergravity in $2+1$ dimensions.
The (restricted) $\osp_L(1|2)$ model, using solely time-like representations,
describes time-like boundaries with fermions on these boundaries.
One should define properly this model, but also see if one could
extend supersymmetrically the continuous series of representations
of $\su(1,1)$ in order to include space-like edges in the model.

It could be also interesting to look at theories with higher
supersymmetries, e.g.\ invariant under $\OSp(p|2)\times
\OSp(q|2)$ \cite{sugra}. As long as the theory can be defined as a
kind of BF theory there does not seem to be any obstacle.

One of the most intriguing further directions is certainly the
generalization to higher dimensions. We have performed the
quantization of super BF theory in any dimension. The interesting
question is thus how (and which) supergravity theories can be
described as constraint super BF theories. The approach would then be
as in popular models of four-dimensional quantum gravity
(for reviews, see \cite{dan:review,alej:review}).
Namely, one tries to implement the constraints directly in
the quantized theory, turning quantum super BF theory into quantum
supergravity.
A delicate issue in this context is
to understand how to write supergravities as gauge theories of
supersymmetically extended Lorentz groups when supersymmetry is an
extension 
of the Poincar{\'e} group. (There is also the problem of giving a
Poincar{\'e} invariant formulation for even dimensional theories.)
It looks as if this direction might finally allow for a connection
between spin foam quantum gravity and superstring theory.
If feasible, this could
provide the starting point for an approach to M-theory that is
both manifestly background inependent and non-perturbative.

Many aspects of the present work need further investigation.
For instance, one should check the asymptotics of
the supersymmetric $6j$-symbols and see if one can recover a Regge action
for gravity+fermions. Extending this to the context of a cosmological
constant, one should compute the
$6j$-symbol for the q-deformed $\OSp_q(1|2)$ and also compute
its asymptotics in order to check the relation between $q$ and the
cosmological constant $\Lambda$.

The super-Ponzano-Regge model as well as the higher dimensional
quantum super
BF theories provide invariants of manifolds that are potentially
new. They might be related to the Casson invariant as is
suggested by Witten's description of such a relation for similar
supersymmetric theories \cite{Witten:CS}.
However, at present this issue is not
clear to us.

\section*{Acknowledgements}

E.~L.\ would like to thank Yi Ling and Daniele Oriti, with whom he had
started the project of studying three-dimensional supergravity, for many discussions
on this subject. R.~O.\ thanks John Barrett for discussions. He
was supported by a Marie Curie Fellowship of the European Union.

\appendix

\section{Partition Functions for the Ponzano-Regge model}

\subsection{The original model: the graded partition function}

We define the Racah symbol as a recoupling coefficient
(with $\Delta_j=2j+1$ the dimension of the
$j$-representation of $\su(2)$):

\bes
&&|(j_1,(j_2,j_3)j_{23})j\ra= \nonumber \\
&&\sum_{j_{12}} (\Delta_{j_{12}}\Delta_{j_{23}})^{1/2}
(-1)^{j_1+j_2+j_3+j}
\left\{\begin{array}{ccc}
j_1 & j_2 & j_{12} \\
j_3 & j & j_{23}
\end{array}\right\}_{R}
|((j_1,j_2)j_{12},j_3)j\ra\nonumber,
\ees
which leads to the expression in terms of the Clebsh-Gordan coefficients
\bes
\left\{\begin{array}{ccc}
j_1 & j_2 & j_{12} \\
j_3 & j & j_{23}
\end{array}\right\}_{R}&=
(-1)^{sign}
\sum_{m_1,m_2,m_3} &
\left[\begin{array}{ccc}
j_1 & j_2 & j_{12} \\
m_1 & m_2 & m_{12}
\end{array}\right] 
\left[\begin{array}{ccc}
j_{12} & j_3 & j \\
m_{12} & m_3 & m
\end{array}\right] \nonumber \\ &&
\left[\begin{array}{ccc}
j_2 & j_3 & j_{23} \\
m_2 & m_3 & m_{23}
\end{array}\right]^*
\left[\begin{array}{ccc}
j_1 & j_{23} & j \\
m_1 & m_{23} & m
\end{array}\right]^*\nonumber,
\ees
with $sign=j_1+j_2+j_3+j$.
Then using the (Biedenharn-Elliot) identity
\bes
&&\left\{\begin{array}{ccc}
g & h & j \\
e & a & d
\end{array}\right\}_{R}
\left\{\begin{array}{ccc}
g & h & j \\
e' & a' & d'
\end{array}\right\}_{R}= \nonumber \\
&&\sum_x \Delta_x (-1)^{sign}
\left\{\begin{array}{ccc}
a & a' & x \\
d' & d & g
\end{array}\right\}_{R}
\left\{\begin{array}{ccc}
d & d' & x \\
e' & e & h
\end{array}\right\}_{R}
\left\{\begin{array}{ccc}
e & e' & x \\
a' & a & j
\end{array}\right\}_{R} \nonumber
\ees
with $sign=g+h+j+e+a+d+e'+a'+d'+x$,
and the corresponding $1\leftrightarrow 4$ identity, one can easily show
that the following model is topological:
\be
Z_{PR}=\sum_{\{j\}}\prod_{edges}(-1)^{2j}\Delta_j
\prod_{triang}(-1)^{a+b+c}
\prod_{tetra} \{6j\}_{R}.
\ee

\subsection{The standard ungraded model}

One can use the Wigner 3j-symbols:
$$
\left(\begin{array}{ccc}
j_1 & j_2 & j_3 \\
m_1 & m_2 & m_3
\end{array}\right)=
\f{(-1)^{j_1-j_2-m_3}}{\sqrt{\Delta_{j_3}}}
\left[\begin{array}{ccc}
j_1 & j_2 & j_3 \\
m_1 & m_2 & -m_3
\end{array}\right]
$$
to define the 6j-symbol:
\bes
\left\{\begin{array}{ccc}
j_1 & j_2 & j_3 \\
j_4 & j_5 & j_6
\end{array}\right\}&=
\sum_{m_i}(-1)^{sign} &
\left(\begin{array}{ccc}
j_1 & j_2 & j_3 \\
m_1 & m_2 & m_3
\end{array}\right)
\left(\begin{array}{ccc}
j_5 & j_6 & j_1 \\
m_5 & -m_6 & m_1
\end{array}\right) \nonumber \\ &&
\left(\begin{array}{ccc}
j_6 & j_4 & j_2 \\
m_6 & -m_4 & m_2
\end{array}\right)
\left(\begin{array}{ccc}
j_4 & j_5 & j_3 \\
m_4 & -m_5 & m_3
\end{array}\right) \nonumber
\ees
with the $sign=j_4+j_5+j_6+m_4+m_5+m_6$ due to using
the dual representation for $j_4,j_5,j_6$.
The partition function for the Ponzano-Regge model is then simply:
\be
Z_{PR}=\sum_{\{j\}}\prod_{edges}\Delta_j
\prod_{tetra} \{6j\}.
\ee

If one writes the partition function for a closed 3-manifold made out
of two identical tetrahedra glued together, one can check that
\be
\{6j\}_R^2=\{6j\}^2,
\ee
so the difference boils down to a difference in signs.




\end{document}